\documentclass[sigconf,screen,authorversion]{acmart}

\usepackage{array}

\usepackage{adjustbox}
\usepackage{xr}
\usepackage{hyperref}
\graphicspath{{figs/}{figures/}{pictures/}{images/}{./}} 

\usepackage{multirow}
\usepackage[table,xcdraw]{xcolor}

\usepackage{tabu}                      
\usepackage{booktabs}                  
\usepackage{lipsum}                    
\usepackage{mwe}                       
\usepackage{amsmath}                   
\usepackage{graphicx}    

\usepackage{subcaption}
\usepackage{placeins}
\usepackage{xcolor}

\usepackage[normalem]{ulem}
\usepackage{amsmath}

\usepackage{submission}
\usepackage{macros}

\newcommand{\update}[1]{\textcolor{black}{#1}}
\newcommand{\mup}[1]{\textcolor{black}{#1}}
\newcommand{\updateNEW}[1]{\textcolor{black}{#1}}

\setmode{submission}

\setcopyright{acmlicensed}
\copyrightyear{2026}
\acmYear{2026}
\acmDOI{XXXXXXX.XXXXXXX}

\acmConference[GI '26]{GI 2026: Graphics Interface}{June 9--12, 2026}{Waterloo, Canada}
\acmBooktitle{GI 2026: Graphics Interface,
  June 9--12, 2026, Waterloo, Canada}
\acmISBN{978-1-4503-XXXX-X/18/06}

\begin{document}



\title{Comparing Controller-Free Pointing Techniques Across Depth for 2D Selection in Augmented Reality}

\author{Samiha Sultana}
\orcid{0009-0005-5780-9614}
\affiliation{%
  \institution{Carleton University}
   \city{Ottawa}
  \country{Canada}}
\email{SamihaSultana@cmail.carleton.ca}

\author{J. Felipe Gonzalez}
\orcid{0000-0002-0716-1689}
\affiliation{%
  \institution{Carleton University}
   \city{Ottawa}
  \country{Canada}}
\email{johannavila@cunet.carleton.ca}

\author{Robert J. Teather}
\orcid{0009-0007-4572-1820}
\affiliation{%
 \institution{Monash University}
   \city{Melbourne}
  \country{Australia}}
\email{rob.teather@monash.edu}

\renewcommand{\shortauthors}{Sultana et al.}

\begin{abstract}
This paper presents a systematic evaluation of five controller-free pointing techniques for 2D target selection in AR, using ISO 9241-411. We compared them across multiple depths (2\,m, 6\,m, 10\,m) in terms of movement time, accuracy, throughput, and workload \mbox{(NASA TLX)}. Head- and eye-based pointing significantly outperformed the hand-based methods (Finger, Wrist, and Arm); Head input was the most accurate and remained the most consistent across depth. 
Depth significantly impacted performance, with complex interactions with target size and distance.
Our results offer a comprehensive empirical basis for selecting appropriate controller-free techniques in depth-varying AR tasks.

\end{abstract}



\hyphenpenalty=9500  
\tolerance=1000
\emergencystretch=1em


\begin{teaserfigure}
 \centering
  \includegraphics[width=0.9\textwidth]{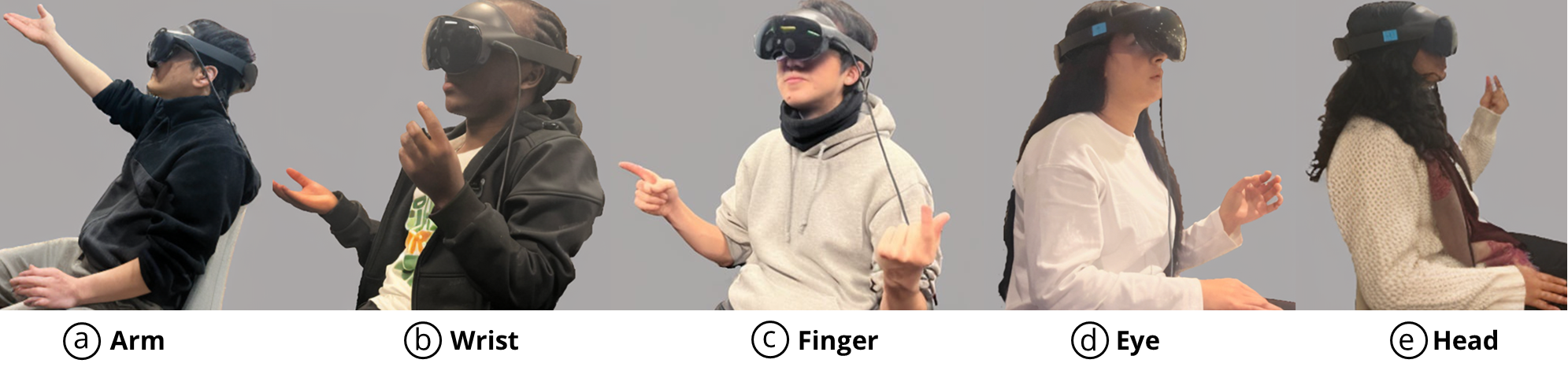}
 \caption{The five controller-free pointing techniques for study: (a) Arm, (b) Wrist, (c) Finger, (d) Eye, and (e) Head. Each was tested at 2, 6, and 10 meters of depth to evaluate movement time, error rate, throughput, and workload in AR 2D selection.}
	\label{fig:tech}
\end{teaserfigure}


\maketitle

\section{Introduction}

Remote pointing facilitates interaction in both augmented reality (AR) and virtual reality (VR), allowing users to aim at distant points or select and manipulate objects beyond physical reach~\cite{argelaguet_efficient_2009, wagner_fitts_2023, mine_virtual_1995}. Traditionally, remote pointing requires controllers to manipulate the origin and direction of a virtual ray; pressing a button afterward completes the interactions, such as confirming object selection or locomotion~\cite{kopper_human_2010, bowman_testbed_1999}. 
However, recent advances in hand-~\cite{schafer_controlling_2021, chowdhury_wriarm_2022}, head-, and eye-tracking~\cite{kyto_pinpointing_2018,pfeuffer_gaze_2017} support new \textbf{controller-free} pointing techniques including arm-, wrist-, finger-, head-, and eye-based methods. These present competitive alternatives, potentially eliminating the need for physical controllers~\cite{buckingham_hand_2021}.

Prior studies have explored the benefits and limitations of these techniques. 
Hand-based input (e.g., arm, wrist, and finger) is intuitive and supports proprioception~\cite{lin_investigation_2015, bernardos_comparison_2016}, but it can yield fatigue~\cite{hincapie-ramos_consumed_2014, czerwinski_toward_2003}. Wrist-based pointing is less fatiguing~\cite{chowdhury_wriarm_2022}, but also less precise, especially for very far targets~\cite{myers_interacting_2002, nancel_high-precision_2013}. The finger is widely used for manipulation, gestures, and UI interaction~\cite{mangalam_enhancing_2024, kim_atatouch_2021, bowman_pinch_2001}, but due to tracking limitations it is rarely used for remote pointing and is underexplored~\cite{oakley_pointing_2008}. Head-based pointing offers stable, hands-free interaction~\cite{lee_comparison_2023}, while eye-gaze is fast~\cite{kyto_pinpointing_2018, pfeuffer_gaze_2017}, but it is sensitive to calibration errors and involuntary eye movements~\cite{blattgerste_advantages_2018}. 
\felipe{I miss a phrase here saying why comparing them and evaluating them is important.} \sam{added}

Unfortunately, prior work differs widely in focus and methodology. 
There are evaluations of specific combinations of pointing technique and selection indication mechanisms, such as gaze with dwell, pinch, or button click~\cite{mutasim_pinch_2021, blattgerste_advantages_2018}. Other work evaluated only limited subsets of modalities, such as gaze, foot, and head~\cite{minakata_pointing_2019}, or compared controllers with gaze or finger input~\cite{fernandes_looking_2025, mifsud_augmented_2022, kim_atatouch_2021}. 
Several studies explore compound interactions (e.g., gaze with hand input), but do not isolate the individual contributions of each modality~\cite{wagner_eye-hand_2024, sidenmark_eye_2019, lystbaek_gaze-hand_2022}. 
Most prior research evaluates an input modality \emph{relative to} a hand-held controller~\cite{cournia_gaze-_2003, sidenmark_comparing_2023, fernandes_looking_2025, ocampo2025comparing}, rather than directly comparing controller-free modalities against each another.
To date there are no systematic evaluations isolating hand-, finger-, or wrist-based pointing performance~\cite{yu_object_2024}. \mup{Comparing these input modalities (i.e., finger, wrist, arm, eye and head) systematically is important because each one offers distinct strengths and limitations (e.g., movement time, error rate, and workload) that are often confounded in techniques that combine multiple input modalities.}

Another difficulty in comparing previous results lies in the variety of study tasks, such as menu selection in AR~\cite{blattgerste_advantages_2018} or in virtual and real-world environments~\cite{zhao_movement_2023}. 
ISO 9241-411~\cite{noauthor_isots_2012} is based on Fitts’ law~\cite{fitts_information_1954}, and improves study comparability through use of a standardized task and measures. Fitts' law is a robust predictive model of pointing performance and is widely applied in VR~\cite{huang_design_2019, schafer_controlling_2022, prithul_evaluation_2022, amini_systematic_2025} and, to a lesser extent, in AR~\cite{cao_real-time_2023}.
According to Fitts’ law, pointing performance is influenced by target distance and width in both VR~\cite{minakata_pointing_2019, wagner_eye-hand_2024, mutasim_pinch_2021, clark_extending_2020, kim_effect_2025} and AR~\cite{zhao_movement_2023, kyto_pinpointing_2018, batmaz_head-mounted_2019}. 
Most Fitts’ law evaluations in 3D environments 
overlook the influence of target depth in 3D selection tasks~\cite{amini_systematic_2025}.
Yet, in 3D environments target depth influences pointing performance due to both perspective scaling and since the ray's angular precision degrades with distance~\cite{kovacs_perceptual_2008, hourcade_how_2012, kopper_human_2010}. Stereo displays, including all commercially available head-mounted displays (e.g., Meta Quest Pro), are also subject to vergence-accommodation conflicts, impacting depth perception and compromising interaction~\cite{westermeier_assessing_2024, batmaz_head-mounted_2019}. 

We present a study addressing these gaps via a rigorous, systematic, and generalizable evaluation. 
\updateNEW{Most prior pointing research focused on VR~\cite{huang_design_2019, khundam_first_2015, schafer_controlling_2022, zhang_double_2017, franzluebbers_versatile_2023, prithul_evaluation_2022}, with comparatively fewer studies in AR~\cite{cao_real-time_2023, naceri_depth_2010, jones_effects_2008}. Although early work reported differences between VR and AR~\cite{jones_effects_2008, naceri_depth_2010}, more recent findings suggest that this gap may be smaller with improved hardware~\cite{batmaz_head-mounted_2019}. We therefore conduct an up-to-date, systematic comparison of controller-free pointing techniques in AR.}
Our research question is: \textbf{\textit{What are the performance and user experience differences between controller-free remote pointing techniques in AR?}}

We employed the ISO 9241-411~\cite{noauthor_isots_2012} standard 
selection task to compare accuracy and completion time of five controller-free remote pointing techniques: Finger, Wrist, Arm, Eye, and Head (Figure \ref {fig:tech}). 
Additionally, we assessed user workload using the NASA-TLX questionnaire~\cite{hart_nasa-task_2006}.
By comparing these interaction modalities, we offer insights into which techniques are best suited to AR tasks with varying spatial demands.

In summary, the main contributions of our work include:

\begin{itemize}
    \item A systematic evaluation of controller-free pointing techniques (Finger, Wrist, Arm, Eye, and Head) in AR.
    \item A thorough examination of the impact of target depth on selection performance with these techniques.
    \item An analysis of pointing efficiency and the speed/accuracy trade-off across input modalities using Fitts’ law metrics.
\end{itemize}

\section{Background}

\subsection{Pointing in AR/VR}

Remote pointing supports reaching specific points or objects beyond arm's reach~\cite{kopper_human_2010, janzen_modeling_2016, shoemaker_two-part_2012} in both VR and AR~\cite{argelaguet_efficient_2009, laviola_3d_2017}. It supports distal interaction while reducing the need for locomotion~\cite{xu_pointing_2019}.
It involves visually identifying a target, \textbf{pointing} with a cursor (e.g., via ray-casting), and \textbf{confirming} selection through actions like a button press, a pinch gesture, or dwell~\cite{mutasim_pinch_2021, blattgerste_advantages_2018, bowman_testbed_1999}. 

Prior work suggests that using separate hands for pointing and confirmation improves performance: Sindhupathiraja et al.~\cite{sindhupathiraja_exploring_2024} reported better results when users pointed with the dominant hand and confirmed with a non-dominant-hand pinch than when the same hand was used for both actions. This division helps mitigate the Heisenberg effect, where executing the confirmation action can degrade pointing accuracy~\cite{wolf2020understanding, narbayev2025exploring}. Consequently, all confirmations in our study used a non-dominant-hand pinch gesture.

While remote pointing historically relied on controllers, modern devices (e.g., Meta Quest 3~\cite{meta_meta_2025}) enable \textit{controller-free} pointing through head-, hand-, and eye-tracking, supporting modalities such as Head, Eye, Arm, Wrist, and Finger~\cite{schafer_controlling_2021, chowdhury_wriarm_2022, kyto_pinpointing_2018, pfeuffer_gaze_2017}. 
Prior work examined multimodal combinations (e.g., eye+head or eye+hand) and confirmation mechanisms (pinch, click, gestures, dwell)~\cite{sidenmark_eyehead_2019, lystbaek_gaze-hand_2022, wagner_fitts_2023, pfeuffer_gaze_2017, mutasim_pinch_2021, lin_investigation_2015}. Other studies focused on aiming dynamics (e.g., gaze behavior or target properties)~\cite{sidenmark_eye_2019, clark_extending_2020, mifsud_augmented_2022}. However, these studies usually emphasize combinations or dynamics rather than fundamental comparisons of standalone pointing techniques.

Several studies have directly compared gaze and head pointing against other inputs. Minakata \ea~\cite{minakata_pointing_2019} compared gaze, head, and foot pointing and reported head outperforming gaze. Lin \ea~\cite{lin_investigation_2015} found hand better for tracking, while gaze was more effective for discrete pointing. Hansen \ea~\cite{hansen_fitts_2018} compared mouse, head, and gaze, and Qian and Teather~\cite{qian_eyes_2017} reported head-based pointing outperforming eye and the combination of eye and head.

In contrast, Arm-, Wrist-, and Finger-based pointing remains underexplored as standalone selection techniques. Although common for navigation or gesture interaction in VR~\cite{huang_design_2019, khundam_first_2015, schafer_controlling_2022, zhang_double_2017}, they have rarely been evaluated in isolation. Early work explored finger/wrist/arm using wearables~\cite{oakley_pointing_2008}, but not in VR and is less representative of current AR tracking. Studies using hand-held controllers~\cite{cournia_gaze-_2003, sidenmark_comparing_2023, fernandes_looking_2025} differ fundamentally from controller-free modalities. A survey of 106 papers on VR/AR selection and manipulation~\cite{yu_object_2024} reported no direct comparisons of these hand-based modalities as standalone selection techniques.

Moreover, most pointing research targets VR~\cite{huang_design_2019, khundam_first_2015, schafer_controlling_2022, zhang_double_2017, franzluebbers_versatile_2023, prithul_evaluation_2022}, with fewer studies in AR~\cite{cao_real-time_2023, naceri_depth_2010, jones_effects_2008}. While early work reported differences between VR and AR~\cite{jones_effects_2008, naceri_depth_2010}, more recent findings suggest smaller gaps, potentially due to improved hardware~\cite{batmaz_head-mounted_2019}. This motivates an up-to-date, systematic comparison of controller-free pointing techniques in AR.

\subsection{Selection and Fitts' Law}

Target selection techniques, including remote pointing, are \update{commonly} evaluated using ISO 9241-411~\cite{noauthor_isots_2012}, and Fitts' law~\cite{hansen_fitts_2018,mackenzie_fitts_1992,oakley_pointing_2008,monteiro_evaluation_2023}. Fitts' law predicts movement time (MT) to acquire a target as:

\begin{equation}
  \begin{split}
MT=a+b \log _{2}\left(\frac{A}{W}+1\right)\\ 
    \end{split}  
    \label{eqn_Fittslaw_popular}
\end{equation}
  \rule{1.0em}{0.0em} 
where \emph{A} is the movement amplitude (i.e., distance to the target), \emph{W} the target width (\ie, target size). The log term is the index of difficulty, or \emph{ID}, which increases with higher amplitudes and/or smaller targets. The coefficients \emph{a} and \emph{b} are empirically derived through linear regression between recorded average movement time and $ID$. 
To mitigate speed-accuracy trade-off biases \cite{zhai_speedaccuracy_2004}, an accuracy adjustment \cite{crossman_speed_1957} is applied to calculate the effective Index of Difficulty, $ID_e$. 
Using $ID_e$, throughput ($TP$) is calculated as: 
\begin{equation}
\begin{split}
TP = \frac{ID_e}{MT} , \quad \text{where} \quad ID_e = \log_2\left(\frac{A_e}{W_e} + 1\right)
\end{split}  
\label{throughput-formula}
\end{equation}

Here, effective width, $W_e = 4.133 \cdot SD_{x}$ where $SD_{x}$ is the standard deviation of distance between selection coordinates and the target centre. This accounts for variability in selection coordinates. $W_e$ is an adjusted target size where 96\% of selections would have hit the target (i.e., 4\% error rate). This accuracy adjustment facilitates comparing throughput across studies with varying error rates \cite{zhai_speedaccuracy_2004}. Effective amplitude, $A_e$, is the average distance from the cursor start position to the end position in each trial. Both effective measures better capture the task participants perform, rather than what is presented.
We used Fitts' law to analyze pointing effectiveness to improve comparability to previous research.

\subsection{Depth Perception}

Perceptual problems persist in AR/VR displays, impacting target acquisition. Notably, the convergence/accommodation mismatch impacts user ability to localize targets in 3D~\cite{hoffman_vergenceaccommodation_2008, fernandes_looking_2025}. Target depth --- the distance of the interaction plane from the user~\cite{tao_freehand_2021, janzen_modeling_2016, shi_pointing_2022} --- also influences pointing performance~\cite{kovacs_perceptual_2008, hourcade_how_2012, kopper_human_2010}. Due to perspective, target depth directly relates to target width in display space (i.e., farther targets appear smaller), which in turn affects selection time~\cite{kovacs_perceptual_2008, hourcade_how_2012, kim_effect_2025}. 
Teather and Stuerzlinger~\cite{teather_pointing_2013} found that stereo cursors can help with near-screen targets but degrade performance for deeper ones due to diplopia (i.e., double vision). 
\updateNEW{Recent work~\cite{bashar2025effect} in VR using modern headsets like the Quest 3 confirmed that selection performance significantly degrades (i.e., increasing movement time and reducing throughput) as targets move away from the focal plane due to the vergence-accommodation conflict.}
Triantafyllidis and Li~\cite{triantafyllidis_challenges_2021} highlight existing research gaps in VR/AR pointing, which notably includes that the impact of depth perception issues is \update{underexplored}. 
\update{We therefore evaluate performance across multiple target depths for several controller-free pointing modalities.}

\section{Pointing Technique Design Space}
Pointing in AR is influenced by choice of pointing technique, target depth, and, per Fitts' law, target size and distance. We discuss the contributions of each below.

\subsection{Pointing Techniques}
\label{sec:pointing-tech}

\mup{We selected five controller-free pointing techniques to cover the main tracking capabilities of modern AR HMDs. This included three hand-based ray origins (Finger, Wrist, Arm) alongside two hands-free ray origins (Head, Eye). This facilitates controlled evaluation of how different ray origins affect selection performance under identical conditions (e.g., confirmation and cursor).}

\begin{figure}[htbp]
	\centering
	\includegraphics[ width=1.0\columnwidth]{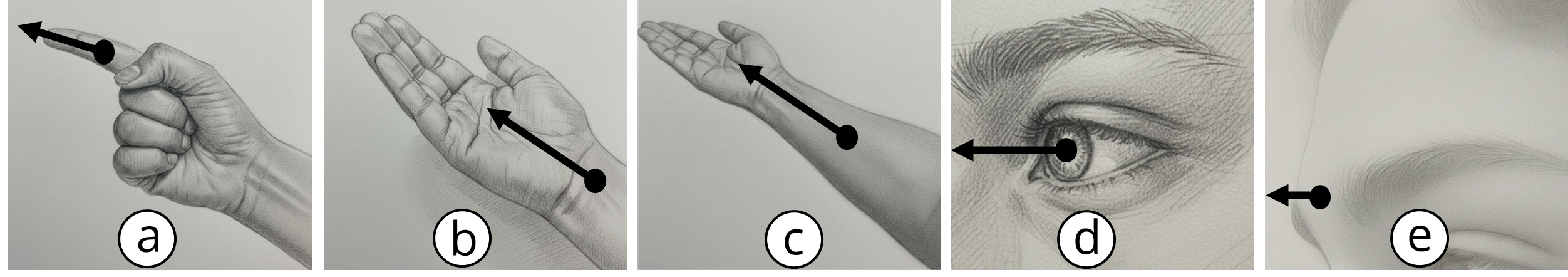}
	\caption{The five pointing techniques, depicting ray origin and direction: (a) Finger, (b) Wrist, (c) Arm, (d) Eye, (e) Head.}
   
	\label{fig:AR} 
\end{figure}

\textbf{\emph{1. Finger:}} By aligning the index finger with the target (Figure \ref {fig:AR}a), participants rely on very fine motor control in the finger~\cite{oakley_pointing_2008}.
The selection ray is cast from the fingertip, allowing rapid pointing.

\textbf{\emph{2. Wrist:}} The pointer ray originates from the user’s dominant wrist (Figure \ref {fig:AR}b), leveraging fine motor control at the wrist joint \cite{chowdhury_wriarm_2022, chowdhury_paws_2023}.
The device’s built-in SDK\footnote{\href{https://developer.oculus.com/documentation/unity/unity-isdk-interaction-sdk-overview/}{https://developer.oculus.com/documentation/unity/unity-isdk-interaction-sdk-overview/}} handles tracking, and wrist orientation controls the ray’s direction. \updateNEW{Wrist was included because the ray direction is directly steered by wrist rotation, allowing fine-grained control through the wrist joint itself.}

\textbf{\emph{3. Arm:}} The pointer ray originates from the distal forearm (Figure \ref {fig:AR}c) and the ray direction is controlled by the user's arm orientation. This relies on gross motor control with the ray anchored at the lower arm region \cite{chowdhury_wriarm_2022}.  \updateNEW{We included both Arm and Wrist because they rely on different control mechanisms. In Arm, the ray is primarily steered by forearm/arm orientation, and wrist rotation does not affect cursor movement.}

\textbf{\emph{4. Eye:}} Targets were selected via gaze using the device eye tracker\footnotemark[1], see Figure \ref {fig:AR}d. 
The ray direction was determined by the eye gaze vector provided by the device SDK\footnotemark[1], which represents the user’s gaze in world space based on the tracked eye pose.

\textbf{\emph{5. Head:}} The pointer ray originates from the midpoint between the user’s eyes (Figure~\ref {fig:AR}e), following the approach in Movement SDK\footnote{\href{https://developer.oculus.com/documentation/unity/move-overview/}{https://developer.oculus.com/documentation/unity/move-overview/}}.
This frees the hands~\cite{kyto_pinpointing_2018} and is useful when arms are fatigued or engaged in other tasks \cite{myers_interacting_2002}.

No filtering (e.g., smoothing) was applied to any of the five techniques. This ensured that our evaluation captures the baseline performance of each method without the confounding influence of a filtering algorithm \cite{manakhov_filtering_2024} or the latency they can introduce \cite{teather_effects_2009}.


\begin{figure*}[htbp]
	\centering
	\includegraphics[width=0.9\textwidth]{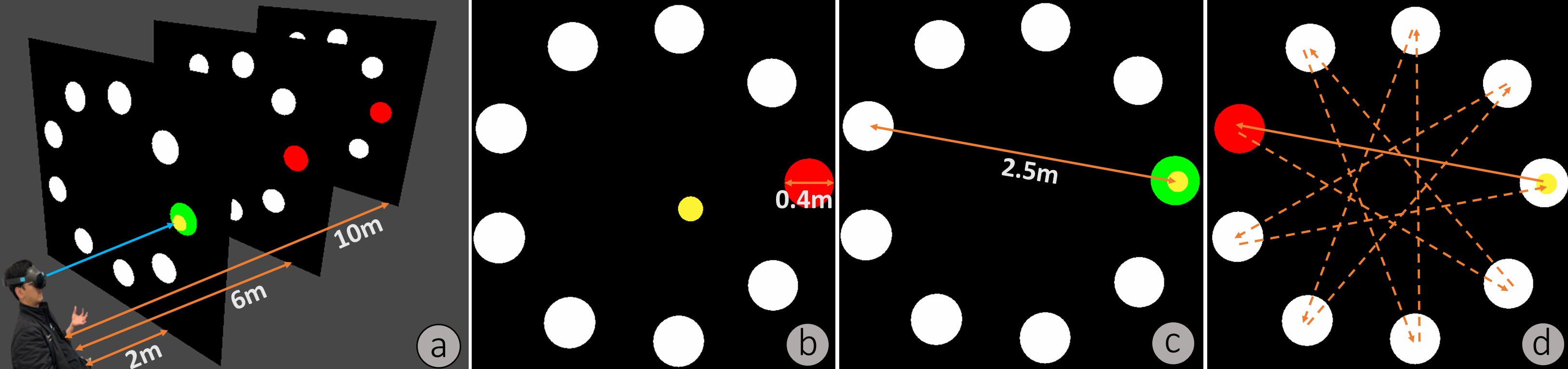}
 \caption{ (a) Study setup with different target depth conditions, (b) yellow cursor approaching target with 0.4m target width, (c) yellow cursor colliding with target with green visual cue and the target amplitude is 2.5m, and (d) After confirming the selection the next target turns red. The solid arrow shows the path to the current target, while dashed arrows represent the sequential order in which the user must select the 9 targets.}
 \label{task}
\end{figure*} 

\subsection{Target Width and Amplitude}

According to Fitts' law, target size and distance influence selection task difficulty (\emph{ID}) and thus target acquisition time. It is thus customary to vary target width (\emph{W}) and movement amplitude (\emph{A}) to present a wide range of IDs, in accordance with ISO 9241-411 \cite{soukoreff_towards_2004, noauthor_isots_2012}. This provides the advantage of comparing different pointing techniques/devices across a range of task difficulties (e.g., easy selection tasks vs. hard selection tasks), enhancing experimental external validity.
Previous work mostly considered linear target width and movement amplitude, i.e., measured in meters or pixels \cite{fitts_information_1954, tao_freehand_2021, lischke_magic-pointing_2016, haque_myopoint_2015, siddhpuria_pointing_2018, nancel_high-precision_2013}. Therefore, we used target width of 0.2\,m, 0.3\,m, and 0.4\,m and for target amplitude we used 1.5\,m, 2.5\,m, and 3.5\,m. The combination of these width and amplitude values results in 9 distinct \emph{ID} levels, ranging from 2.24 to 4.21 bits (see eqn. 1). \updateNEW{Gori et al.~\cite{gori2018perils} cautioned that confounding effects in Fitts' law studies can arise when different amplitude/width combinations yield the same \emph{ID}, especially in designs with geometrically distributed amplitudes and widths. In our study, amplitude and width values were not geometrically distributed, and the nine amplitude and width combinations yielded nine distinct \emph{ID} values.}


\subsection{Target Depth}
In our study, target depth is the perpendicular distance from the user to the interaction plane where the targets reside \cite{tao_freehand_2021, janzen_modeling_2016, shi_pointing_2022}. There are known issues relating to depth perception \cite{triantafyllidis_challenges_2021, hoffman_vergenceaccommodation_2008}, especially across different input methods \cite{triantafyllidis_challenges_2021}. We therefore investigate 3 depth values (2\,m, 6\,m, and 10\,m away from the viewer) across the five pointing techniques.  
\mup{We maintained constant target widths in meters across depths (similar to previous Fitts' law studies \cite{barrera_machuca_effect_2019, clark_extending_2020, sindhupathiraja_exploring_2024, teather_pointing_2013}) rather than fixed angular widths. While this design decision inherently varies visual angle, it prioritizes ecological validity: in real-world AR scenarios, physical objects do not dynamically scale with user distance.}
\updateNEW{Table \ref{tab:angularwidths} summarizes the corresponding visual angles for each amplitude/width/depth combination.}

\begin{table}[bh]

\scriptsize
\begin{tabular}{|p{1.2cm}l|lllllllll|}
\hline
\multicolumn{2}{|l|}{}                                                & \multicolumn{9}{c|}{\textbf{Depth (m)}}                                                                                                                                                                                                                                                                                                                                                                                                                           \\ \cline{3-11} 
\multicolumn{2}{|l|}{\multirow{-2}{*}{\textbf{}}}                     & \multicolumn{3}{c|}{\cellcolor[HTML]{FFFFFF}\textbf{2}}                                                                                                    & \multicolumn{3}{c|}{\cellcolor[HTML]{EFEFEF}\textbf{6}}                                                                                                   & \multicolumn{3}{c|}{\cellcolor[HTML]{FFFFFF}\textbf{10}}                                                                             \\ \hline
\multicolumn{2}{|c|}{\textbf{Amplitude (m)}}                               & \multicolumn{1}{l|}{\cellcolor[HTML]{FFFFFF}1.5}   & \multicolumn{1}{l|}{\cellcolor[HTML]{FFFFFF}2.5}  & \multicolumn{1}{l|}{\cellcolor[HTML]{FFFFFF}3.5}  & \multicolumn{1}{l|}{\cellcolor[HTML]{EFEFEF}1.5}  & \multicolumn{1}{l|}{\cellcolor[HTML]{EFEFEF}2.5}  & \multicolumn{1}{l|}{\cellcolor[HTML]{EFEFEF}3.5}  & \multicolumn{1}{l|}{\cellcolor[HTML]{FFFFFF}1.5}  & \multicolumn{1}{l|}{\cellcolor[HTML]{FFFFFF}2.5}  & \cellcolor[HTML]{FFFFFF}3.5  \\ \hline
\multicolumn{1}{|c|}{}                                 & \textbf{0.2} & \multicolumn{1}{l|}{\cellcolor[HTML]{FFFFFF}5.36}  & \multicolumn{1}{l|}{\cellcolor[HTML]{FFFFFF}4.86} & \multicolumn{1}{l|}{\cellcolor[HTML]{FFFFFF}4.31} & \multicolumn{1}{l|}{\cellcolor[HTML]{EFEFEF}1.89} & \multicolumn{1}{l|}{\cellcolor[HTML]{EFEFEF}1.87} & \multicolumn{1}{l|}{\cellcolor[HTML]{EFEFEF}1.83} & \multicolumn{1}{l|}{\cellcolor[HTML]{FFFFFF}1.14} & \multicolumn{1}{l|}{\cellcolor[HTML]{FFFFFF}1.14} & \cellcolor[HTML]{FFFFFF}1.13 \\ \cline{2-11} 
\multicolumn{1}{|c|}{}                                 & \textbf{0.3} & \multicolumn{1}{l|}{\cellcolor[HTML]{FFFFFF}8.03}  & \multicolumn{1}{l|}{\cellcolor[HTML]{FFFFFF}7.28} & \multicolumn{1}{l|}{\cellcolor[HTML]{FFFFFF}6.46} & \multicolumn{1}{l|}{\cellcolor[HTML]{EFEFEF}2.84} & \multicolumn{1}{l|}{\cellcolor[HTML]{EFEFEF}2.80} & \multicolumn{1}{l|}{\cellcolor[HTML]{EFEFEF}2.75} & \multicolumn{1}{l|}{\cellcolor[HTML]{FFFFFF}1.71} & \multicolumn{1}{l|}{\cellcolor[HTML]{FFFFFF}1.71} & \cellcolor[HTML]{FFFFFF}1.69 \\ \cline{2-11} 
\multicolumn{1}{|c|}{\multirow{-3}{*}{\raisebox{0.1\height}[3.5ex][1.75ex]{\rotatebox[origin=c]{90}{\parbox{1.2cm}{\centering\textbf{Width \\(m)}}}}}} & \textbf{0.4} & \multicolumn{1}{l|}{\cellcolor[HTML]{FFFFFF}10.70} & \multicolumn{1}{l|}{\cellcolor[HTML]{FFFFFF}9.70} & \multicolumn{1}{l|}{\cellcolor[HTML]{FFFFFF}8.61} & \multicolumn{1}{l|}{\cellcolor[HTML]{EFEFEF}3.79} & \multicolumn{1}{l|}{\cellcolor[HTML]{EFEFEF}3.74} & \multicolumn{1}{l|}{\cellcolor[HTML]{EFEFEF}3.67} & \multicolumn{1}{l|}{\cellcolor[HTML]{FFFFFF}2.29} & \multicolumn{1}{l|}{\cellcolor[HTML]{FFFFFF}2.27} & \cellcolor[HTML]{FFFFFF}2.26 \\ \hline
\end{tabular}

\caption{\updateNEW{Angular widths in degrees for each combination of width (W), amplitude (A), and depth (D).}}
\label{tab:angularwidths}
\end{table}

\vspace{-3em}
\section{Methodology}

We conducted a systematic evaluation of five controller-free pointing techniques for AR target selection, evaluating both target depth and selection task difficulty.



    

\subsection{Participants}

We recruited 20 participants (11 men, 9 women), aged 19 to 48 years (mean = 27.95, \emph{SD} = 8.51), from the local university. All participants were right-handed, had (corrected-to-) normal vision, with no known motor impairments. Of the 20 participants, 8 had no prior experience with AR/VR. \updateNEW{The study was conducted in Canada; each participant provided consent in accordance with Carleton University's Research Ethics Board protocol, and received \$15 CAD for their participation.} 

\subsection{Apparatus}

We used a Meta Quest Pro headset for its built-in hand- and eye-tracking capabilities, high-resolution (1800 x 1920 pixels per eye) displays, and wide field of view (106° horizontal, 96° vertical). 
Participants sat in a quiet laboratory setting, free from distractions, ensuring consistent experimental conditions.

\begin{figure}[htbp]
	\centering
	\includegraphics[trim= 1cm 1cm 1cm 3cm, clip, width=0.6\columnwidth]{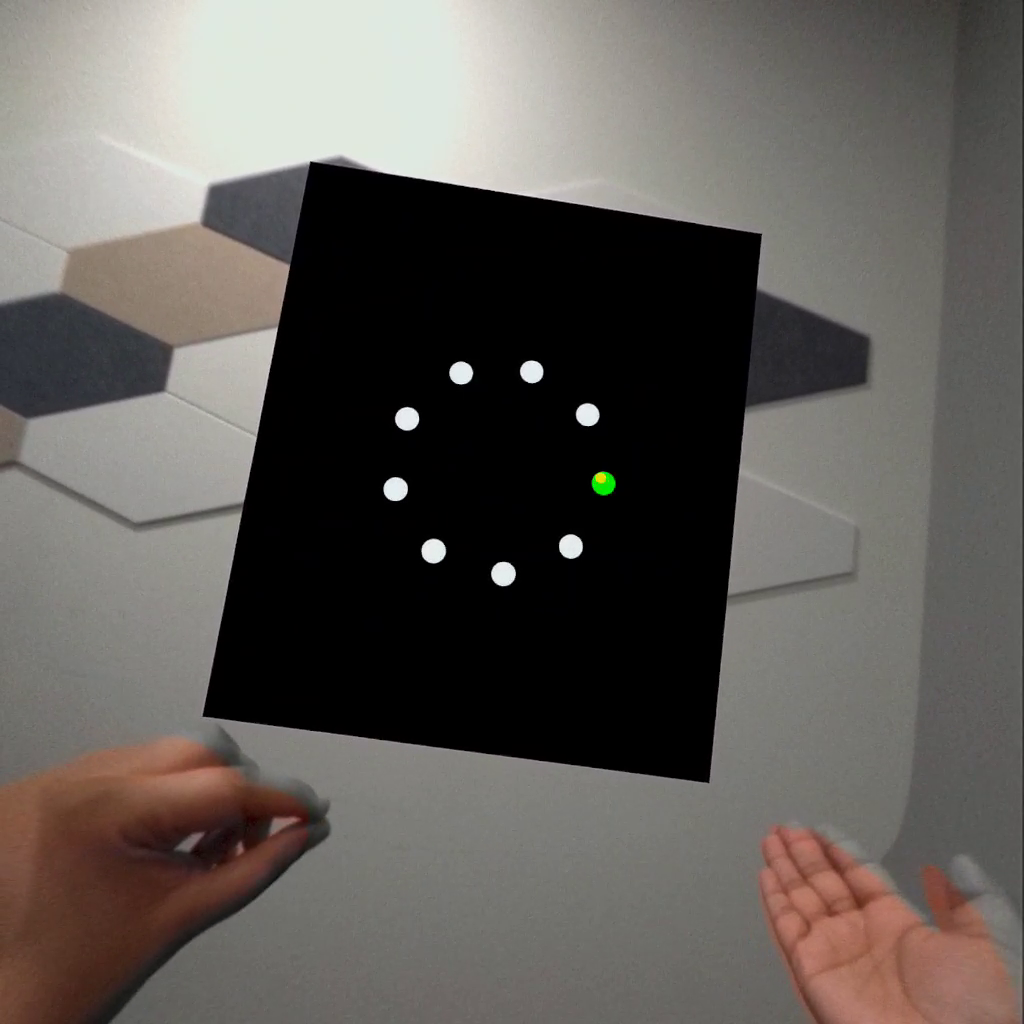}
	\caption{Augmented Reality View of the task from participant's perspective for Wrist. When the yellow cursor intersects the red target (seen in figure \ref{task}), the target turns green to provide visual feedback that it is ready for selection.}
   
	\label{fig:AR-pic} 
\end{figure}

\update{We implemented an ISO 9241-411 reciprocal target selection task in Unity3D. Circular 2D targets were arranged in a ring at three depths (2\,m, 6\,m, 10\,m) with widths (0.2\,m, 0.3\,m, 0.4\,m) and amplitudes (1.5\,m, 2.5\,m, 3.5\,m), yielding nine $ID$ levels (2.24--4.21 bits). The software presented a 20\,cm yellow cursor (as seen in Figure \ref{task} and \ref{fig:AR-pic}), which resided in the plane of the targets where the selection ray intersected that plane. For each trial, the current target changed red while the other targets remained white. When the cursor intersected the current target, it turned green to indicate that it was ready for selection. Selection was confirmed with a pinch gesture by the non-dominant-hand thumb and forefinger. If the cursor was outside the target, the system registered confirmations as a miss; however, the trial continued until the participant successfully selected the target, after which the next target in the sequence would become active and turn red. The target sequence is seen in Figure \ref{task}d. We used Meta’s OVRHand SDK  to obtain tracked poses for the pointing techniques (see section \ref{sec:pointing-tech}) and logged all trial data via Firebase Realtime Database\footnote{\href{https://firebase.google.com/products/realtime-database}{https://firebase.google.com/products/realtime-database}}.}

\subsection{Procedure}

Upon arrival, participants provided written informed consent prior to participating. They were then introduced to the Quest Pro headset and the general study objectives. 
Before starting the actual trials for each pointing technique, participants completed at least 10 practice trials to familiarize themselves with the pointing technique and the confirmation gesture. 
\mup{For each condition, we instructed participants to primarily use the intended pointing technique to control the ray via the specified joint or pose.}
\mup{If they were observed using a joint other than the target joint during the Arm, Wrist, and Finger conditions, we gently reminded them to only use the specified joint.}
\mup{Eye-tracking calibration was performed prior to the eye pointing condition. Head motion was not constrained, as users naturally rotate their heads during target acquisition \cite{nancel_high-precision_2013}, particularly when target amplitude is high \cite{kyto_pinpointing_2018}.}

In each trial, a target appeared at a specified depth, width, and amplitude combination. Participants were instructed to select the target as quickly and accurately as possible using the currently assigned pointing technique. After acquiring the target (i.e., aligning the pointer with the target), they performed a pinch gesture with their non-dominant hand to confirm selection. 

Participants completed a NASA TLX questionnaire and provided comments after each technique.
They were allowed to rest and briefly remove the headset before starting the next technique if needed.
Upon completing all trials with each selection technique, participants completed a short demographic survey and ranked the techniques. We then thanked them for their time and compensated them at the end of the experiment.

\subsection{Design}

Our study followed a 5 × 3 × 3 × 3 within-subjects design with the following independent variables and levels:
\begin{itemize}
    \item \emph{Pointing Technique:} Finger, Wrist, Arm, Eye, Head
    \item \emph{Width:} 0.2\,m, 0.3\,m, 0.4\,m
    \item \emph{Amplitude:} 1.5\,m, 2.5\,m, 3.5\,m
    \item \emph{Depth:} 2\,m, 6\,m, 10\,m
\end{itemize}
 
\updateNEW{For each combination of independent variables, the user had to perform 9 trials.} Each trial involved selecting a single target. Pointing technique order was counterbalanced using a Latin square design to mitigate order effects. The 9 combinations of Amplitude and Width yielded 9 distinct \emph{ID}s, representing varying task difficulty. For each Pointing Technique, the order of \emph{ID} and Depth were randomized.
Each participant completed 5 (techniques) × 9 (IDs) × 3 (depths) × 9 (trials per ID) = 1215 trials, for a total of 24,300 trials across all 20 participants.

Our dependent variables included: 

\begin{itemize}
    \item \textbf{Movement Time:} The duration between selecting the previous target and selecting the current target, in seconds.
   
    \item \textbf{Error rate:} The percentage of trials where the selection ended outside the target boundary (i.e., missed the target).
  
    \item  \textbf{NASA TLX:} Subjective workload ratings via the NASA TLX questionnaire. \updateNEW{We calculated the overall workload using raw NASA-TLX values.}

    \item \textbf{Throughput:} We also calculate the \textbf{Throughput (TP)} derived from movement time and effective \emph{ID} (via effective width and amplitude) using equation \ref{throughput-formula}.

\end{itemize}

\section{Results}


We removed outliers where the movement time exceeded three standard deviations from the mean (1.65\% of data), leaving 23,898 trials. We analyzed time and throughput via repeated measures ANOVA with Bonferroni corrections. Sphericity violations were corrected using Greenhouse-Geisser ($\epsilon_{GG}<.75$) or Huynh-Feldt ($\epsilon_{GG}\ge.75$). Error rate and NASA-TLX data were analyzed using the Aligned Rank Transform (ART) ANOVA \cite{wobbrock_aligned_2011, elkin_aligned_2021}.

\begin{figure}[tphb!]
  \small
  \includegraphics[width=1.0\columnwidth]{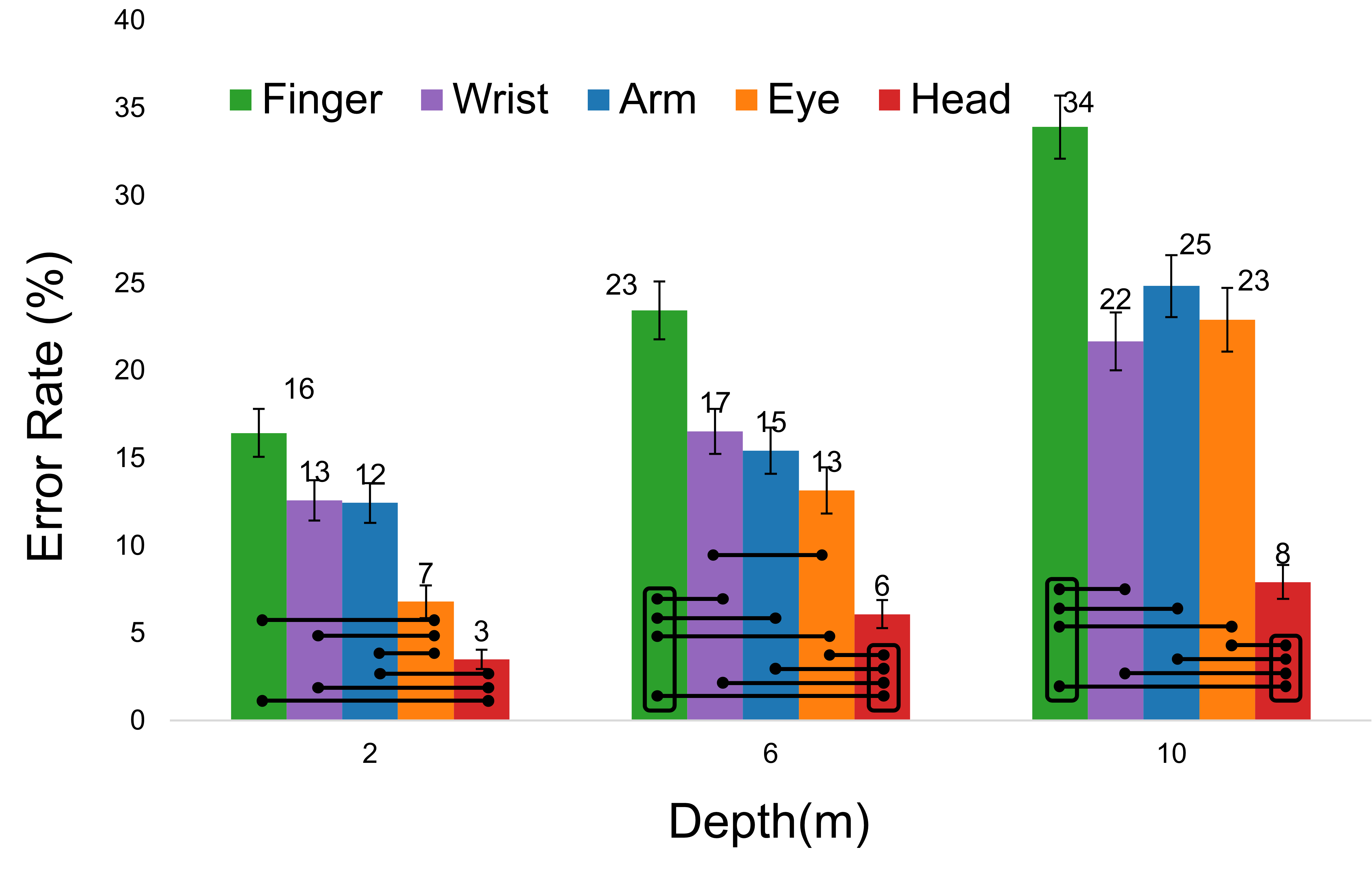}
  \caption{\updateNEW{Error Rate per Technique and Depth. Error bars indicate standard error.}}
  \label{fig:err-combined}
\end{figure}

In Figures \ref{fig:movement-time}, \ref{fig:err-combined}, and \ref{fig:tp-bar}, pairwise significant differences are depicted as horizontal lines where \emph{p} < .05. Squares enclosing the endpoints of these lines indicate conditions that were significantly different from all other conditions. Significant differences across depths are omitted for clarity.

\subsection{Movement Time}

\begin{figure}[htbp!]
  \centering
    \includegraphics[width=0.8\linewidth]{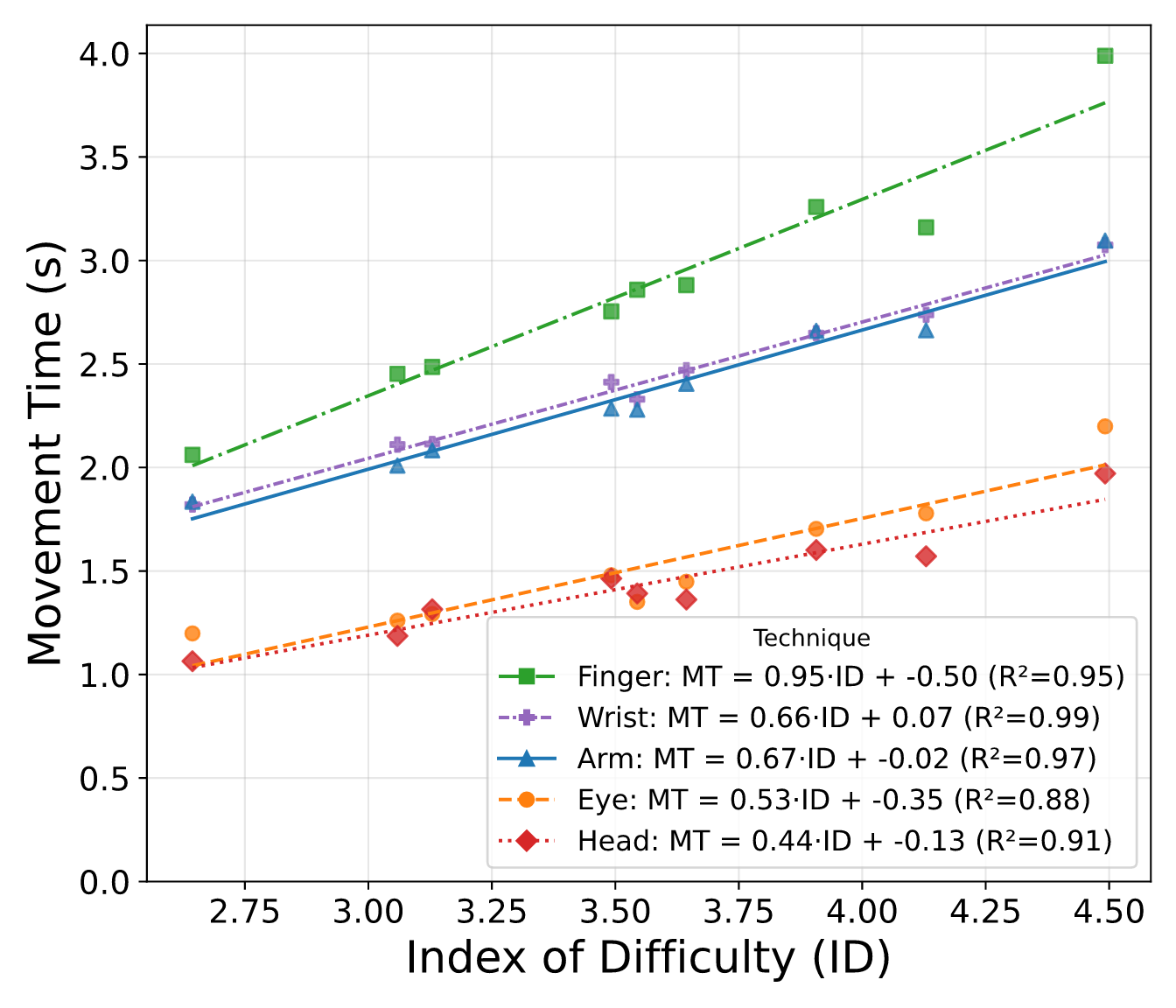}
  \caption{Movement Time vs Index of Difficulty }
  \label{fig:fitts-mt}
\end{figure}

We found that Fitts' law accurately models movement time in terms of \emph{ID}, with $R^2 \geq 0.88$ for all techniques, as indicated by the Fitts' Law regression lines in Figure \ref{fig:fitts-mt}. The Head- and Eye-based pointing techniques offered the fastest movement times and the smallest sensitivity to \emph{ID}, resulting in faster and more stable performance compared to Arm, Wrist, and Finger techniques.
There were significant main effects for both pointing technique ($F_{4,76}=57.73$, $p<.001$, \textbf{$\eta^2$} $= 0.75$) and depth ($F_{2,38}=63.19$, $p<.001$, \textbf{$\eta^2$} $= 0.77$) on movement time.

Pairwise comparisons of the pointing techniques revealed that both Head (mean = $1.43$\,s, \sd = $0.07$\,s) and Eye (mean = $1.55$\,s, \sd = $0.10$\,s) were significantly faster than Arm (mean = $2.38$\,s, \sd = $0.10$\,s), Wrist (mean = $2.43$\,s, \sd = $0.15$\,s), and Finger (mean = $2.93$\,s, \sd = $0.14$\,s), with all comparisons significant at $p < .001$. 
For \textit{Depth}, all pairwise comparisons between all three levels—2\,m (mean = $1.91$\,s, \sd = $0.08$\,s), 6\,m (mean = $2.03$\,s, \sd = $0.08$\,s), and 10\,m (mean = $2.50$\,s, \sd = $0.12$\,s) were significantly different. 

\begin{figure}[htbp!]
 	\centering
         \small
 	\includegraphics[width=0.95\columnwidth]{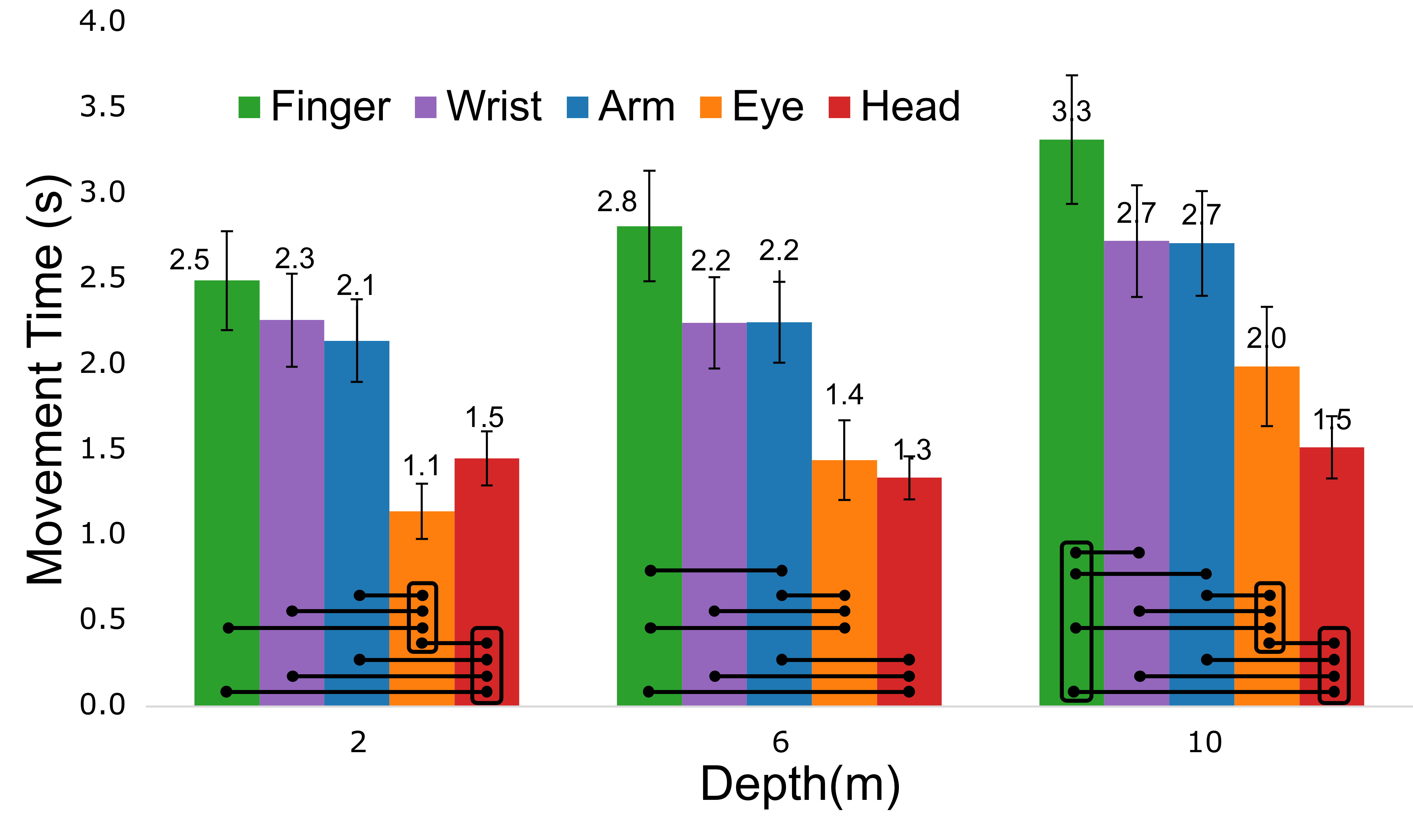}
  \caption{ \updateNEW{Mean Movement Time per depth per technique.
Error bars indicate standard error.}}
  \label{fig:movement-time}
 \end{figure}

The interaction effect Pointing Technique $\times$ Depth was also significant ($F_{4.06,152}=6.97$, $p<.001$, \textbf{$\eta^2$} $= 0.27$). Pairwise comparisons revealed depth-related differences between most pointing techniques, with a few exceptions: for the Arm, there was no significant difference between 2\,m (mean = 2.15\,s, \sd = 0.13\,s) and 6\,m (mean = 2.25\,s, \sd = 0.10\,s); for the Head, there was no significant difference between 2\,m (mean = 1.45\,s, \sd = 0.07\,s) and 6\,m (mean = 1.34\,s, \sd = 0.06\,s), nor between 2\,m and 10\,m (mean = 1.52\,s, \sd = 0.10\,s). For the Wrist, there was no significant difference between 2\,m (mean = 2.28\,s, \sd = 0.14\,s) and 6\,m (mean = 2.26\,s, \sd = 0.15\,s).
As seen in Figure~\ref{fig:movement-time}, depth effects generally followed the main effect trends, with a few notable deviations. At 2\,m, Eye was the fastest pointing technique (mean = 1.14\,s, \sd = 0.07\,s), significantly outperforming all others, including the Head (mean = 1.45\,s, \sd = 0.07\,s). At 2\,m depth, the Arm and Finger pointing techniques were not significantly different; however, they were significantly different at the 6\,m and 10\,m depths.
At 10\,m depth, the Head pointing technique (mean = 1.50\,s, \sd = 0.10\,s) was fastest, significantly outperforming Eye (mean = 2.00\,s, \sd = 0.16\,s). Notably, performance with Head was stable regardless of depth; MT with all other pointing techniques increased with greater depth.

It is also worth noting that amplitude and width influenced the finger and wrist pointing techniques, yielding significant Pointing Technique $\times$ Amplitude ($F_{3.96,75.29}=3.68$, $p<.01$, \textbf{$\eta^2$} $= 0.16$) and Pointing Technique $\times$ Width ($F_{8,152}=11.39$, $p<.001$, \textbf{$\eta^2$} $= 0.38$) interaction effects. 
Specifically, at an amplitude of 1.5\,m, the performance of Finger and Wrist was not significantly different, while at 2.5\,m and 3.5\,m, Finger was significantly slower than Wrist.
For width, there was no significant difference at the largest width (0.4\,m), but Finger was significantly slower than Wrist at the smaller widths of 0.2\,m and 0.3\,m.



\subsection{Error Rates }


There were significant main effects on error rate for Pointing Technique ($F_{4,2565}=170.92$, $p<.001$) and Depth ($F_{4,2565}=180.21$, $p<.001$). 
Head had a lower error rate \updateNEW{(mean = 5.81\,\%, \sd = 0.70\,\%) } than the other four pointing techniques.

Pairwise comparisons revealed significant differences between all technique pairs, except for Arm and Wrist. 
\updateNEW{Eye (mean = 14.27\,\%, \sd = 1.53\,\%) also had a lower error rate than Wrist (mean = 16.91\,\%, \sd = 1.39\,\%), Arm (mean = 17.55\,\%, \sd = 1.61\,\%), and Finger (mean = 24.58\,\%, \sd = 1.40\,\%).}
Similarly, all depths were significantly different from one another. A significant interaction effect between Pointing Technique and Depth ($F_{8,2565}=9.94$, $p<.001$), \update{indicated Arm error rates were stable between 2 m and 6 m, while Eye and Finger degraded significantly at every depth interval. Head and Wrist remained stable between 6 m and 10 m, suggesting superior stability with greater depth.}

\subsection{Throughput}

\begin{figure}[thpb!]
  \centering
    \includegraphics[width=1.0\linewidth]{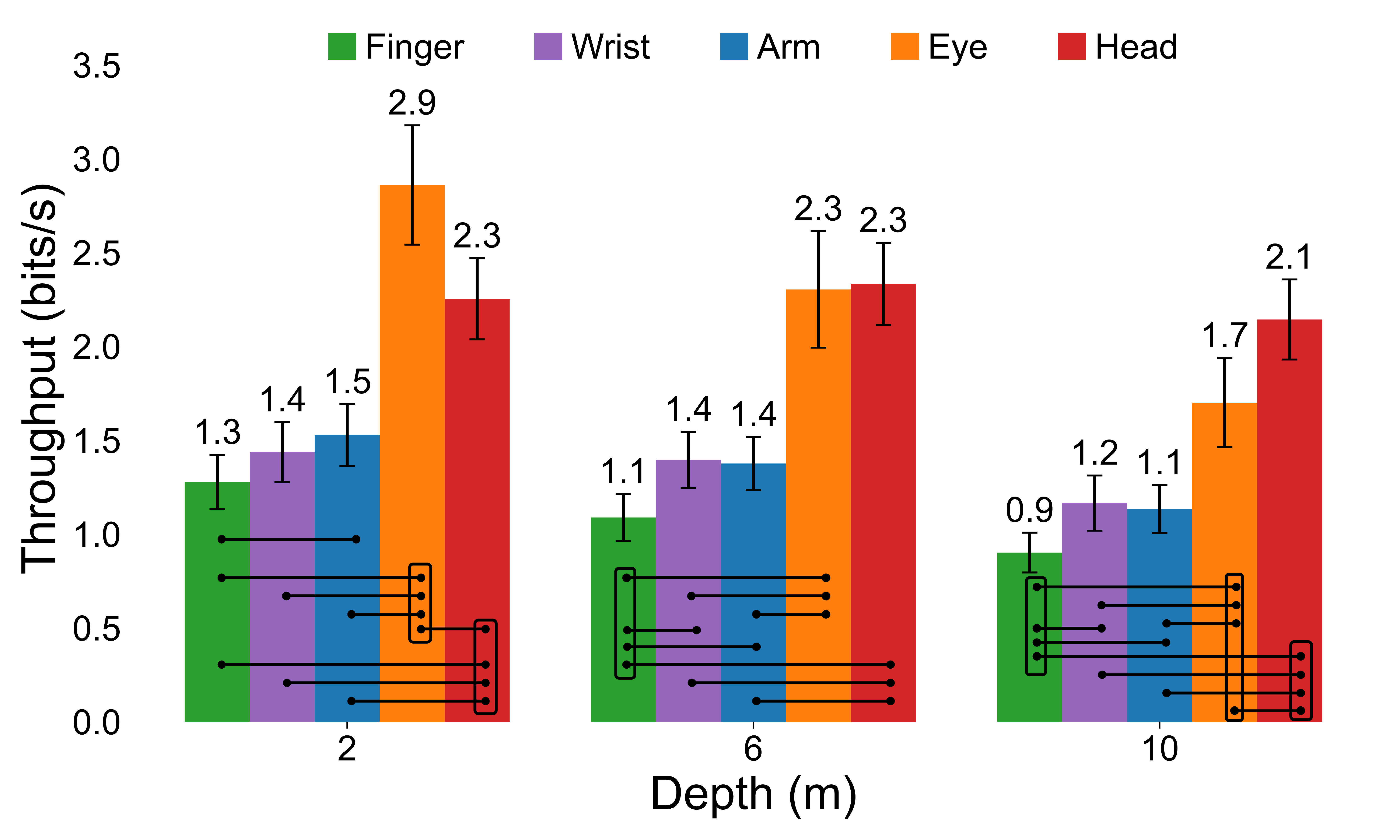}
    \caption{\updateNEW{Throughput per technique per depth. Error bars indicate standard error.}}
  \label{fig:tp-bar}
\end{figure}

There were significant main effects for Pointing Technique ($F_{1.83,34.81}=56.45$, \textbf{$\eta^2$} $= 0.75$) and Depth ($F_{2,38}=125.58$, \textbf{$\eta^2$} $= 0.87$) on throughput. 
Pairwise comparisons between the Pointing Techniques (see Figure \ref{fig:tp-bar}) reveal that Head (mean = $2.41$ bits/s, \sd = 0.11 bits/s) and Eye (mean = $2.53$ bits/s, \sd = 0.16 bits/s) offered significantly higher (all $p < .001$) throughput than Arm (mean = $1.48$ bits/s, \sd = 0.08 bits/s), Wrist (mean = $1.47$ bits/s, \sd = 0.09 bits/s) and Finger (mean = $1.21$ bits/s, \sd = 0.06 bits/s). Additionally, it was found that all other pointing techniques had significantly higher throughput than Finger (all $p<.005$). 

\begin{figure*}[tbhp!]
  \centering

    \includegraphics[width=0.87\linewidth]{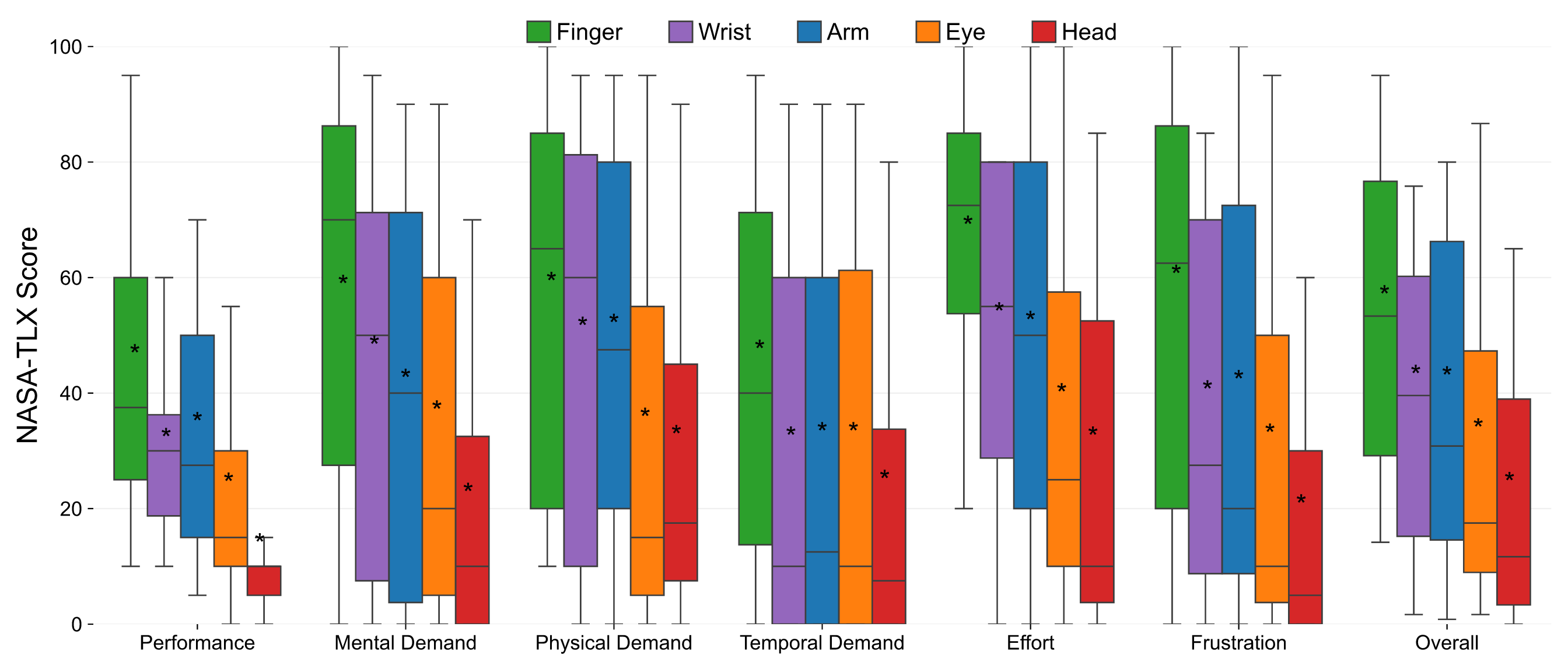}
  

  \caption{NASA TLX scores vs. criteria per Technique.}
  \label{nasatlx}
\end{figure*}

For the effect of Depth, pairwise comparisons revealed that pointing from 2\,m (mean = $2.00$ bits/s, \sd = 0.09 bits/s) had significantly higher (all $p<.001$) $TP$ than both 6\,m (mean = $1.89$ bits/s, \sd = 0.08 bits/s) and 10\,m (mean = $1.57$ bits/s, \sd = 0.08 bits/s). The significant interaction effect between Pointing Technique and Depth presented as for the 2\,m and 6\,m depths, both Eye and Head had higher throughput than Arm, Wrist, and Finger, while at 10\,m, Head had a higher throughput than Eye, Arm, Wrist, and Finger.


\subsection{NASA TLX}




We found significant differences between Pointing Techniques across all NASA TLX metrics (\autoref{nasatlx}): Performance ($F_{4,76}=18.72$, $p<.001$), Mental Demand ($F_{4,76}=14.6$, $p<.001$), Physical Demand ($F_{4,76}=6.59$, $p<.001$), Temporal Demand ($F_{4,76}=6.43$, $p<.001$), Effort ($F_{4,76}=11.5$, $p<.001$), Frustration ($F_{4,76}=11.97$, $p<.001$), and Overall Task Load ($F_{4,76}=16.83$, $p<.001$). 

Post-hoc comparisons revealed that Head was consistently rated as the best technique. It outperformed all other techniques in terms of perceived Performance, and showed significantly lower scores in Mental Demand (vs. all), Effort (vs. Arm, Wrist, and Finger), and Frustration (vs. Arm, Wrist, and Finger). For Physical Demand, Head was significantly lower than Finger, better than Arm, but not significantly different from Eye or Wrist. 

The Eye-based pointing technique also performed well, with higher perceived Performance than Finger, and lower Mental and Physical Demand than Finger. In contrast, the Finger-based pointing technique was consistently rated as the most demanding technique. It received the highest (i.e., worst) scores in Mental and Temporal Demand, Effort, Frustration, and Overall Task Load, being significantly worse than most or all other techniques in each case. 

\vspace{-1em}

\subsection{Pointing Technique Ranking}

\begin{figure}[hb!]
  \centering
  \large
    \includegraphics[width=0.9\linewidth]{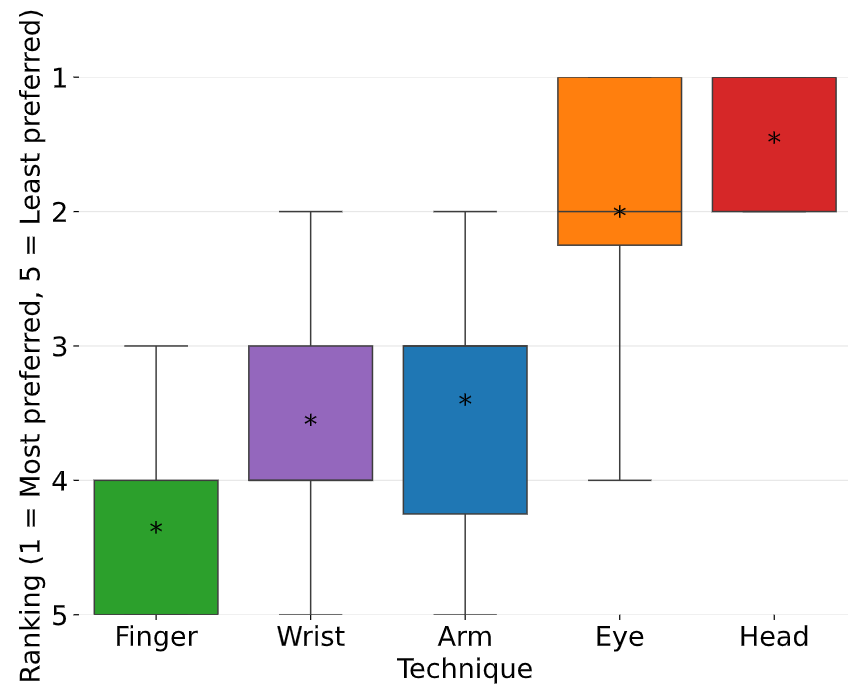}

  \caption{Subjective preference rankings per Technique.}
  \label{fig:ranking-bar}
\end{figure}

Participants ranked the five techniques according to their subjective preference, as illustrated in Figure \ref{fig:ranking-bar}. Head was the most preferred technique (Median = 1, IQR = 1 to 2), with 12 participants ranking it first and 7 ranking it second. Eye also received high preference (Median = 2, IQR = 1 to 2), with 8 participants ranking it first.
Arm was typically mid-ranked (Median = 3, IQR = 3 to 4.25), with 10 participants placing it third, whereas 5 ranked it last, leading to its higher variability.
In contrast to Arm, Wrist showed more stable opinions (Median = 4, IQR = 3 to 4), with most participants consistently ranking it fourth (12 in total).
Finally, Finger was consistently rated as the least preferred technique (Median = 5, IQR = 4 to 5), with 13 participants placing it in the fifth position.

\section{Discussion}

We analyze our findings in terms of both performance and user experience, focusing on differences across pointing techniques and the effects of depth on selection tasks.

\subsection{Pointing Techniques}

Our results reveal that head- and eye-based pointing outperformed all hand-based pointing techniques across all dependent variables. 
Head was 40\% faster than Arm, 41\% faster than Wrist, and 51\% faster than Finger. Eye was 35\% faster than Arm, 36\% faster than Wrist, and 47\% faster than Finger, while Arm and Wrist were 19\% and 17\% faster than Finger, respectively.
The comparable performance between Head and Eye techniques can be due to the coordination of eye and head movements in target acquisition \cite{sidenmark_eyehead_2019}. Users naturally look at a target before or while turning their head towards it, making the eye- and head-based pointing faster compared to the arm, wrist, and finger. While eye-gaze offers faster initial target acquisition \cite{kyto_pinpointing_2018}, its advantage is often offset by higher error rates, particularly at greater depths.
\update{Head had significantly lower error rates than Eye, consistent with prior work \cite{minakata_pointing_2019, qian_eyes_2017, hansen_fitts_2018, kyto_pinpointing_2018}.}
Participants described Head as ``stable'' 
and ``accurate''.
Three rated it as the best or most consistent technique. 
Among hand-based pointing techniques, Arm and Wrist performed similarly, while Finger yielded the worst results.
\update{The similar performance of Arm and Wrist can be due to them being physically linked, so arm motion also shifts the wrist pose \cite{oakley_pointing_2008}.}
Results based on interaction effects show 
more stable error rates with varying target width: 
Finger in small targets, Wrist in large targets, and Arm in both.
This challenges previous assumptions suggesting that Finger technique could offer fast and intuitive control \cite{mifsud_augmented_2022, chowdhury_wriarm_2022}. \update{Four participants described Finger as hard to control and difficult for selection. 
} \updateNEW{One stated, ``Finger [was] difficult to make precise selections.''}
\update{NASA TLX matched performance trends: Head was rated best overall, while Finger was most demanding and frustrating.}

\vspace{-1em}

\subsection{Effect of Depth}

\update{Depth degraded performance for all techniques except Head, increasing movement time and error rates. This is consistent with perspective effects and reduced angular precision at larger distances \cite{kopper_human_2010, kovacs_perceptual_2008, fernandes_looking_2025}.}
\updateNEW{However, some participants noted that Head was sometimes fatiguing, particularly for the target depth 2m, as one 
stated, ``close targets [were] hard on [the] neck but [Head had] high accuracy.''} 
\update{Eye was fastest at close range \cite{pfeuffer_gaze_2017, sidenmark_eyehead_2019}, but its movement time and error rate worsened with distance \cite{wagner_fitts_2023}.}
\updateNEW{One participant 
stated ``closer targets were easier, best for big close targets.''}
Other participants reported that they found Eye was ``fast and easy'' 
, but one 
noted that ``it was difficult to select smaller targets''.
\updateNEW{The pattern may explain why Head outperformed Eye at farther depths: because target visual angle decreased with depth, gaze-based pointing may have been more sensitive than head-based pointing to the reduced apparent target size.}
\update{Arm and Wrist showed similar movement times at 2\,m and 6\,m, suggesting a threshold before depth costs become evident.} 
\updateNEW{Participants found Arm to be worse at farther distances. For example, 
one remarked that ``Arm hurts for smaller and farther targets,'' while another observed that ``Arm was tiring and not stable for long distance.''}
Finger degraded between 2 m and 6 m, suggesting higher sensitivity to depth. Head remained stable across depths, with limited differences in movement time and error rate, making it a strong choice for medium to long-range selection in depth-varying AR.
\update{Movement time and error rates increased with amplitude and decreased with target width, in line with Fitts' law \cite{fitts_information_1954}. But some depth interactions plateaued (e.g., no significant difference of movement time between 2\,m and 6\,m at 2.5--3.5\,m amplitudes, and no significant difference of error rate for 0.3\,m vs 0.4\,m widths at depths 6 m and 10 m). Throughput increased from 1.5 m to 2.5–3.5 m amplitudes, and with larger target widths. Interestingly, 0.2 m targets had significantly lower TP than wider ones, suggesting users may trade speed for precision with narrow targets.}
\update{In summary, our design space analysis reveals a clear trade-off: Head is the most stable and depth-robust \cite{lee_comparison_2023}; Eye is the fastest but depth-sensitive \cite{kyto_pinpointing_2018}; and Hand techniques, while familiar, suffer from fatigue and instability \cite{myers_interacting_2002}.}

\section{Limitations and Future Work}

We conducted the experiment under controlled stationary conditions, using a standardized 2D selection task (using ISO 9241-411) in AR. Participants reported that the solid black background improved focus by removing distractions (P18). While this controlled design enabled fair comparison, future work should examine how locomotion, lighting, and complex 3D AR environments affect performance and workload for these pointing techniques.
\updateNEW{The absence of a hand-held controller baseline is a limitation. While our aim was to compare controller-free techniques in isolation, including a standard controller-based pointing condition would have provided a useful reference for interpreting their relative performance, which future studies could explore.}
\updateNEW{Another limitation is that target widths were kept constant in meters across depths, which caused visual angle to decrease as depth increased. Although this reflects many real-world AR scenarios, the reduced apparent target size at farther depths may have contributed to poorer performance for some techniques, particularly Eye, on smaller targets.}
\updateNEW{Finally, our findings were derived using a Meta Quest Pro headset, which supports video see-through augmented reality. Earlier work reported differences between AR and VR \cite{jones_effects_2008, naceri_depth_2010}, while some more recent studies found smaller or non-significant differences between the two platforms~\cite{batmaz_head-mounted_2019}. Since our task used controller-free selection on 2D panels, we believe the relative differences between techniques found in our environment (video see-through AR) may also apply to VR and optical see-through AR. However, future work should compare video see-through AR, optical see-through AR, and VR directly to confirm the generalizability of our results.}


\section{Conclusion}

We presented a systematic evaluation of five controller-free pointing techniques (Finger, Wrist, Arm, Eye, and Head)
for selecting 2D targets at near (2 m), medium (6 m), and far (10 m) depths in an AR environment. By leveraging the Fitts' law, we investigated the impact of target depth along with standard Fitts' law factors (i.e., target width and amplitude) on user performance, focusing on key metrics such as movement time and error rate. Our findings indicate that head-based pointing consistently offers the optimal balance of speed and accuracy across all tested depths, outperforming arm-, wrist-, and finger-based approaches. Eye-based pointing demonstrated reaching the target fastest, but its accuracy was lower compared to head-based pointing. These results suggest that head-based pointing may serve as an effective default technique for a wide range of AR interactions, particularly when precise selection is critical across different depths. Conversely, eye-based pointing could excel in scenarios where rapid target identification is essential but tolerates slight reductions in accuracy. By identifying the strengths and weaknesses of each controller-free pointing technique, our study offers practical guidance to designers and developers of AR systems and contributes to the growing body of knowledge in human-computer interaction.


\bibliographystyle{ACM-Reference-Format}
\bibliography{main.bib}

@inproceedings{gori2018perils,
  title={The perils of confounding factors: how Fitts' law experiments can lead to false conclusions},
  author={Gori, Julien and Rioul, Olivier and Guiard, Yves and Beaudouin-Lafon, Michel},
  booktitle={Proceedings of the 2018 CHI Conference on Human Factors in Computing Systems},
  pages={1--10},
  year={2018},
  doi = {10.1145/3173574.3173770}
}

@article{bashar2025effect,
  title={The effect of visual depth on the vergence--accommodation conflict on 3D selection performance within virtual reality headsets: MR Bashar et al.},
  author={Bashar, Mohammad Raihanul and Barrera Machuca, Mayra Donaji and Stuerzlinger, Wolfgang and Batmaz, Anil Ufuk},
  journal={The Visual Computer},
  volume={41},
  number={12},
  pages={9645--9661},
  year={2025},
  publisher={Springer},
  doi= {10.1007/s00371-025-03990-x}
}

@inproceedings{ocampo2025comparing,
  title={Comparing Hand and Controller Avatars with Hand Tracking and Controller-Based Interaction},
  author={Ocampo, Natalia and Gonzalez, J Felipe and Teather, Robert J},
  booktitle={2025 IEEE International Symposium on Mixed and Augmented Reality (ISMAR)},
  pages={164--174},
  year={2025},
  organization={IEEE},
  doi = {10.1109/ISMAR67309.2025.00029}
}

@inproceedings{wolf2020understanding,
    author = {Wolf, Dennis and Gugenheimer, Jan and Combosch, Marco and Rukzio, Enrico},
    title = {{Understanding the Heisenberg Effect of Spatial Interaction: A Selection Induced Error for Spatially Tracked Input Devices}},
    year = {2020},
    isbn = {9781450367080},
    publisher = {Association for Computing Machinery},
    address = {New York, NY, USA},
    url = {https://doi.org/10.1145/3313831.3376876},
    doi = {10.1145/3313831.3376876},
    booktitle = {Proceedings of the 2020 CHI Conference on Human Factors in Computing Systems},
    pages = {1–10},
    numpages = {10},
    location = {Honolulu, HI, USA},
    series = {CHI '20}
}

@INPROCEEDINGS {narbayev2025exploring,
author = { Narbayev, Bakdauren and Ullah, A K M Amanat and Sin, Jaisie and Lasserre, Patricia and Hasan, Khalad },
booktitle = { 2025 IEEE International Symposium on Mixed and Augmented Reality (ISMAR) },
title = {{ Exploring Pointing and Confirmation Techniques for Teleportation Across Varying Elevations in Virtual Reality }},
year = {2025},
volume = {},
ISSN = {},
pages = {1671-1681},
doi = {10.1109/ISMAR67309.2025.00170},
url = {https://doi.ieeecomputersociety.org/10.1109/ISMAR67309.2025.00170},
publisher = {IEEE Computer Society},
address = {Los Alamitos, CA, USA},
month =Oct}

@article{manakhov_filtering_2024,
	title = {Filtering on the {Go}: {Effect} of {Filters} on {Gaze} {Pointing} {Accuracy} {During} {Physical} {Locomotion} in {Extended} {Reality}},
	volume = {30},
	issn = {1941-0506},
	shorttitle = {Filtering on the {Go}},
	url = {https://ieeexplore.ieee.org/document/10672561},
	doi = {10.1109/TVCG.2024.3456153},
	number = {11},
	urldate = {2025-09-12},
	journal = {IEEE Transactions on Visualization and Computer Graphics},
	author = {Manakhov, Pavel and Sidenmark, Ludwig and Pfeuffer, Ken and Gellersen, Hans},
	month = nov,
	year = {2024},
	pages = {7234--7244},
}

@inproceedings{teather_effects_2009,
	title = {Effects of tracking technology, latency, and spatial jitter on object movement},
	url = {https://ieeexplore.ieee.org/document/4811204},
	doi = {10.1109/3DUI.2009.4811204},
	urldate = {2025-09-12},
	booktitle = {2009 {IEEE} {Symposium} on {3D} {User} {Interfaces}},
	author = {Teather, Robert J. and Pavlovych, Andriy and Stuerzlinger, Wolfgang and MacKenzie, I. Scott},
	month = mar,
	year = {2009},
	pages = {43--50},
}

@article{fitts_information_1954,
	title = {The information capacity of the human motor system in controlling the amplitude of movement},
	volume = {47},
	issn = {0022-1015},
	doi = {10.1037/h0055392},
    url= {https://doi.org/10.1037/h0055392},
	number = {6},
	journal = {Journal of Experimental Psychology},
	author = {Fitts, Paul M.},
	year = {1954},
	note = {Place: US
Publisher: American Psychological Association},
	pages = {381--391},
}

@inproceedings{wobbrock_aligned_2011,
	address = {New York, NY, USA},
	series = {{CHI} '11},
	title = {The aligned rank transform for nonparametric factorial analyses using only anova procedures},
	isbn = {978-1-4503-0228-9},
	url = {https://dl.acm.org/doi/10.1145/1978942.1978963},
	doi = {10.1145/1978942.1978963},
	urldate = {2025-06-11},
	booktitle = {Proceedings of the {SIGCHI} {Conference} on {Human} {Factors} in {Computing} {Systems}},
	publisher = {Association for Computing Machinery},
	author = {Wobbrock, Jacob O. and Findlater, Leah and Gergle, Darren and Higgins, James J.},
	month = may,
	year = {2011},
	pages = {143--146},
}

@inproceedings{elkin_aligned_2021,
	address = {New York, NY, USA},
	series = {{UIST} '21},
	title = {An {Aligned} {Rank} {Transform} {Procedure} for {Multifactor} {Contrast} {Tests}},
	isbn = {978-1-4503-8635-7},
	url = {https://dl.acm.org/doi/10.1145/3472749.3474784},
	doi = {10.1145/3472749.3474784},
	urldate = {2025-06-11},
	booktitle = {The 34th {Annual} {ACM} {Symposium} on {User} {Interface} {Software} and {Technology}},
	publisher = {Association for Computing Machinery},
	author = {Elkin, Lisa A. and Kay, Matthew and Higgins, James J. and Wobbrock, Jacob O.},
	month = oct,
	year = {2021},
	pages = {754--768},
}

@article{yu_object_2024,
	title = {Object {Selection} and {Manipulation} in {VR} {Headsets}: {Research} {Challenges}, {Solutions}, and {Success} {Measurements}},
	volume = {57},
	issn = {0360-0300},
	shorttitle = {Object {Selection} and {Manipulation} in {VR} {Headsets}},
	url = {https://dl.acm.org/doi/10.1145/3706417},
	doi = {10.1145/3706417},
	number = {4},
	urldate = {2025-07-10},
	journal = {ACM Comput. Surv.},
	author = {Yu, Difeng and Dingler, Tilman and Velloso, Eduardo and Goncalves, Jorge},
	year = {2024},
	pages = {98:1--98:34},
}

@inproceedings{cournia_gaze-_2003,
	address = {Ft. Lauderdale, Florida, USA},
	title = {Gaze- vs. hand-based pointing in virtual environments},
	copyright = {https://www.acm.org/publications/policies/copyright\_policy\#Background},
	url = {http://portal.acm.org/citation.cfm?doid=765891.765982},
	doi = {10.1145/765891.765982},
	urldate = {2025-07-10},
	booktitle = {{CHI} '03 extended abstracts on {Human} factors in computing systems  - {CHI} '03},
	publisher = {ACM Press},
	author = {Cournia, Nathan and Smith, John D. and Duchowski, Andrew T.},
	year = {2003},
	pages = {772},
}

@misc{meta_meta_2025,
	title = {Meta {Quest} {MR}, {VR} {Headsets} \& {Accessories}},
	url = {https://www.meta.com/ca/quest},
	language = {en},
	urldate = {2025-03-07},
	author = {Meta},
	year = {2025},
}

@inproceedings{bowman_testbed_1999,
	address = {New York, NY, USA},
	series = {{VRST} '99},
	title = {Testbed evaluation of virtual environment interaction techniques},
	isbn = {978-1-58113-141-3},
	url = {https://dl.acm.org/doi/10.1145/323663.323667},
	doi = {10.1145/323663.323667},
	urldate = {2025-07-02},
	booktitle = {Proceedings of the {ACM} symposium on {Virtual} reality software and technology},
	publisher = {Association for Computing Machinery},
	author = {Bowman, Doug A. and Johnson, Donald B. and Hodges, Larry F.},
	year = {1999},
	pages = {26--33},
}

@article{buckingham_hand_2021,
	title = {Hand {Tracking} for {Immersive} {Virtual} {Reality}: {Opportunities} and {Challenges}},
	volume = {2},
	issn = {2673-4192},
	shorttitle = {Hand {Tracking} for {Immersive} {Virtual} {Reality}},
	url = {https://www.frontiersin.org/journals/virtual-reality/articles/10.3389/frvir.2021.728461/full},
	doi = {10.3389/frvir.2021.728461},
	language = {English},
	urldate = {2025-02-19},
	journal = {Frontiers in Virtual Reality},
	author = {Buckingham, Gavin},
	month = oct,
	year = {2021},
	note = {Publisher: Frontiers},
}

@inproceedings{minakata_pointing_2019,
	address = {New York, NY, USA},
	series = {{ETRA} '19},
	title = {Pointing by gaze, head, and foot in a head-mounted display},
	isbn = {978-1-4503-6709-7},
	url = {https://dl.acm.org/doi/10.1145/3317956.3318150},
	doi = {10.1145/3317956.3318150},
	urldate = {2025-06-27},
	booktitle = {Proceedings of the 11th {ACM} {Symposium} on {Eye} {Tracking} {Research} \& {Applications}},
	publisher = {Association for Computing Machinery},
	author = {Minakata, Katsumi and Hansen, John Paulin and MacKenzie, I. Scott and Bækgaard, Per and Rajanna, Vijay},
	month = jun,
	year = {2019},
	pages = {1--9},
}

@article{fernandes_looking_2025,
	title = {Looking in {Depth}: {Targeting} by {Eye} and {Controller} {Input} for {Multi}-{Depth} {Target} {Placement}},
	volume = {41},
	issn = {1044-7318},
	shorttitle = {Looking in {Depth}},
	url = {https://doi.org/10.1080/10447318.2024.2401657},
	doi = {10.1080/10447318.2024.2401657},
	number = {13},
	urldate = {2025-06-27},
	journal = {International Journal of Human–Computer Interaction},
	author = {Fernandes, Ajoy S. and , T. Scott, Murdison and and Proulx, Michael J.},
	month = jul,
	year = {2025},
	note = {Publisher: Taylor \& Francis,
	pages = {7952--7967},
}}

@article{sidenmark_eye_2019,
	title = {Eye, {Head} and {Torso} {Coordination} {During} {Gaze} {Shifts} in {Virtual} {Reality}},
	volume = {27},
	issn = {1073-0516},
	url = {https://dl.acm.org/doi/10.1145/3361218},
	doi = {10.1145/3361218},
	number = {1},
	urldate = {2025-06-27},
	journal = {ACM Trans. Comput.-Hum. Interact.},
	author = {Sidenmark, Ludwig and Gellersen, Hans},
	year = {2019},
	pages = {4:1--4:40},
}

@article{lystbaek_gaze-hand_2022,
	title = {Gaze-{Hand} {Alignment}: {Combining} {Eye} {Gaze} and {Mid}-{Air} {Pointing} for {Interacting} with {Menus} in {Augmented} {Reality}},
	volume = {6},
	shorttitle = {Gaze-{Hand} {Alignment}},
	url = {https://dl.acm.org/doi/10.1145/3530886},
	doi = {10.1145/3530886},
	number = {ETRA},
	urldate = {2025-06-27},
	journal = {Proc. ACM Hum.-Comput. Interact.},
	author = {Lystbæk, Mathias N. and Rosenberg, Peter and Pfeuffer, Ken and Grønbæk, Jens Emil and Gellersen, Hans},
	month = may,
	year = {2022},
	pages = {145:1--145:18},
}

@inproceedings{wagner_eye-hand_2024,
	address = {New York, NY, USA},
	series = {{UIST} '24},
	title = {Eye-{Hand} {Movement} of {Objects} in {Near} {Space} {Extended} {Reality}},
	isbn = {979-8-4007-0628-8},
	url = {https://dl.acm.org/doi/10.1145/3654777.3676446},
	doi = {10.1145/3654777.3676446},
	urldate = {2025-06-27},
	booktitle = {Proceedings of the 37th {Annual} {ACM} {Symposium} on {User} {Interface} {Software} and {Technology}},
	publisher = {Association for Computing Machinery},
	author = {Wagner, Uta and Asferg Jacobsen, Andreas and Feuchtner, Tiare and Gellersen, Hans and Pfeuffer, Ken},
	month = oct,
	year = {2024},
	pages = {1--13},
}

@inproceedings{blattgerste_advantages_2018,
	address = {New York, NY, USA},
	series = {{COGAIN} '18},
	title = {Advantages of eye-gaze over head-gaze-based selection in virtual and augmented reality under varying field of views},
	isbn = {978-1-4503-5790-6},
	url = {https://dl.acm.org/doi/10.1145/3206343.3206349},
	doi = {10.1145/3206343.3206349},
	urldate = {2025-03-24},
	booktitle = {Proceedings of the {Workshop} on {Communication} by {Gaze} {Interaction}},
	publisher = {Association for Computing Machinery},
	author = {Blattgerste, Jonas and Renner, Patrick and Pfeiffer, Thies},
	month = jun,
	year = {2018},
	pages = {1--9},
}

@inproceedings{wagner_fitts_2023,
	address = {New York, NY, USA},
	series = {{CHI} '23},
	title = {A {Fitts}’ {Law} {Study} of {Gaze}-{Hand} {Alignment} for {Selection} in {3D} {User} {Interfaces}},
	isbn = {978-1-4503-9421-5},
	url = {https://dl.acm.org/doi/10.1145/3544548.3581423},
	doi = {10.1145/3544548.3581423},
	urldate = {2025-06-27},
	booktitle = {Proceedings of the 2023 {CHI} {Conference} on {Human} {Factors} in {Computing} {Systems}},
	publisher = {Association for Computing Machinery},
	author = {Wagner, Uta and Lystbæk, Mathias N. and Manakhov, Pavel and Grønbæk, Jens Emil Sloth and Pfeuffer, Ken and Gellersen, Hans},
	year = {2023},
	pages = {1--15},
}

@inproceedings{sidenmark_eyehead_2019,
	address = {New York, NY, USA},
	series = {{UIST} '19},
	title = {Eye\&{Head}: {Synergetic} {Eye} and {Head} {Movement} for {Gaze} {Pointing} and {Selection}},
	isbn = {978-1-4503-6816-2},
	shorttitle = {Eye\&{Head}},
	url = {https://dl.acm.org/doi/10.1145/3332165.3347921},
	doi = {10.1145/3332165.3347921},
	urldate = {2025-07-01},
	booktitle = {Proceedings of the 32nd {Annual} {ACM} {Symposium} on {User} {Interface} {Software} and {Technology}},
	publisher = {Association for Computing Machinery},
	author = {Sidenmark, Ludwig and Gellersen, Hans},
	month = oct,
	year = {2019},
	pages = {1161--1174},
}

@inproceedings{kim_effect_2025,
	address = {New York, NY, USA},
	series = {{CHI} {EA} '25},
	title = {The {Effect} of {Target} {Depth} on {Performance} of {Multi}-directional {Tapping} {Task} in {Virtual} {Reality}},
	isbn = {979-8-4007-1395-8},
	url = {https://dl.acm.org/doi/10.1145/3706599.3719893},
	doi = {10.1145/3706599.3719893},
	urldate = {2025-07-01},
	booktitle = {Proceedings of the {Extended} {Abstracts} of the {CHI} {Conference} on {Human} {Factors} in {Computing} {Systems}},
	publisher = {Association for Computing Machinery},
	author = {Kim, Haejun and Hong, Yuhwa and Yu, Jihae and Xiong, Shuping and Kim, Woojoo},
	year = {2025},
	pages = {1--8},
}

@inproceedings{amini_systematic_2025,
	address = {New York, NY, USA},
	series = {{CHI} '25},
	title = {A {Systematic} {Review} of {Fitts}' {Law} in {3D} {Extended} {Reality}},
	isbn = {979-8-4007-1394-1},
	url = {https://dl.acm.org/doi/10.1145/3706598.3713623},
	doi = {10.1145/3706598.3713623},
	urldate = {2025-07-01},
	booktitle = {Proceedings of the 2025 {CHI} {Conference} on {Human} {Factors} in {Computing} {Systems}},
	publisher = {Association for Computing Machinery},
	author = {Amini, Mohammadreza and Stuerzlinger, Wolfgang and Teather, Robert J and Batmaz, Anil Ufuk},
	year = {2025},
	pages = {1--25},
}

@article{sidenmark_comparing_2023,
	title = {Comparing {Gaze}, {Head} and {Controller} {Selection} of {Dynamically} {Revealed} {Targets} in {Head}-{Mounted} {Displays}},
	volume = {29},
	issn = {1077-2626},
	url = {https://doi.org/10.1109/TVCG.2023.3320235},
	doi = {10.1109/TVCG.2023.3320235},
	number = {11},
	urldate = {2025-07-01},
	journal = {IEEE Transactions on Visualization and Computer Graphics},
	author = {Sidenmark, Ludwig and Prummer, Franziska and Newn, Joshua and Gellersen, Hans},
	month = nov,
	year = {2023},
	pages = {4740--4750},
}

@article{zhao_movement_2023,
	title = {Movement {Time} for {Pointing} {Tasks} in {Real} and {Augmented} {Reality} {Environments}},
	volume = {13},
	copyright = {http://creativecommons.org/licenses/by/3.0/},
	issn = {2076-3417},
	url = {https://www.mdpi.com/2076-3417/13/2/788},
	doi = {10.3390/app13020788},
	language = {en},
	number = {2},
	urldate = {2025-07-01},
	journal = {Applied Sciences},
	author = {Zhao, Caijun and Li, Kai Way and Peng, Lu},
	month = jan,
	year = {2023},
	note = {Number: 2
Publisher: Multidisciplinary Digital Publishing Institute},
	pages = {788},
}

@inproceedings{mifsud_augmented_2022,
	title = {Augmented {Reality} {Fitts}' {Law} {Input} {Comparison} {Between} {Touchpad}, {Pointing} {Gesture}, and {Raycast}},
	url = {https://ieeexplore.ieee.org/abstract/document/9757658},
	doi = {10.1109/VRW55335.2022.00146},
	urldate = {2025-07-01},
	booktitle = {2022 {IEEE} {Conference} on {Virtual} {Reality} and {3D} {User} {Interfaces} {Abstracts} and {Workshops} ({VRW})},
	author = {Mifsud, Domenick M. and Williams, Adam S. and Ortega, Francisco and Teather, Robert J.},
	month = mar,
	year = {2022},
	pages = {590--591},
}

@article{clark_extending_2020,
	title = {Extending {Fitts}’ law in three-dimensional virtual environments with current low-cost virtual reality technology},
	volume = {139},
	issn = {1071-5819},
	url = {https://www.sciencedirect.com/science/article/pii/S1071581920300173},
	doi = {10.1016/j.ijhcs.2020.102413},
	urldate = {2025-07-01},
	journal = {International Journal of Human-Computer Studies},
	author = {Clark, Logan D. and Bhagat, Aakash B. and Riggs, Sara L.},
	month = jul,
	year = {2020},
	pages = {102413},
}

@inproceedings{mutasim_pinch_2021,
	address = {New York, NY, USA},
	series = {{ETRA} '21 {Short} {Papers}},
	title = {Pinch, {Click}, or {Dwell}: {Comparing} {Different} {Selection} {Techniques} for {Eye}-{Gaze}-{Based} {Pointing} in {Virtual} {Reality}},
	isbn = {978-1-4503-8345-5},
	shorttitle = {Pinch, {Click}, or {Dwell}},
	url = {https://dl.acm.org/doi/10.1145/3448018.3457998},
	doi = {10.1145/3448018.3457998},
	urldate = {2025-07-01},
	booktitle = {{ACM} {Symposium} on {Eye} {Tracking} {Research} and {Applications}},
	publisher = {Association for Computing Machinery},
	author = {Mutasim, Aunnoy K and Batmaz, Anil Ufuk and Stuerzlinger, Wolfgang},
	month = may,
	year = {2021},
	pages = {1--7},
}

@inproceedings{pfeuffer_gaze_2017,
	address = {New York, NY, USA},
	series = {{SUI} '17},
	title = {Gaze + pinch interaction in virtual reality},
	isbn = {978-1-4503-5486-8},
	url = {https://dl.acm.org/doi/10.1145/3131277.3132180},
	doi = {10.1145/3131277.3132180},
	urldate = {2025-07-01},
	booktitle = {Proceedings of the 5th {Symposium} on {Spatial} {User} {Interaction}},
	publisher = {Association for Computing Machinery},
	author = {Pfeuffer, Ken and Mayer, Benedikt and Mardanbegi, Diako and Gellersen, Hans},
	month = oct,
	year = {2017},
	pages = {99--108},
}

@inproceedings{chowdhury_wriarm_2022,
	title = {{WriArm}: {Leveraging} {Wrist} {Movement} to {Design} {Wrist}+{Arm} {Based} {Teleportation} in {VR}},
	shorttitle = {{WriArm}},
	url = {https://ieeexplore.ieee.org/document/9995582},
	doi = {10.1109/ISMAR55827.2022.00047},
	urldate = {2025-04-09},
	booktitle = {2022 {IEEE} {International} {Symposium} on {Mixed} and {Augmented} {Reality} ({ISMAR})},
	author = {Chowdhury, Sohan and Ullah, A K M Amanat and Pelmore, Nathan Bruce and Irani, Pourang and Hasan, Khalad},
	month = oct,
	year = {2022},
	note = {ISSN: 1554-7868},
	pages = {317--325},
}

@article{shoemaker_two-part_2012,
	title = {Two-{Part} {Models} {Capture} the {Impact} of {Gain} on {Pointing} {Performance}},
	volume = {19},
	issn = {1073-0516},
	url = {https://dl.acm.org/doi/10.1145/2395131.2395135},
	doi = {10.1145/2395131.2395135},
	number = {4},
	urldate = {2025-04-07},
	journal = {ACM Trans. Comput.-Hum. Interact.},
	author = {Shoemaker, Garth and Tsukitani, Takayuki and Kitamura, Yoshifumi and Booth, Kellogg S.},
	year = {2012},
	pages = {28:1--28:34},
}

@article{zhai_speedaccuracy_2004,
	series = {Fitts' law 50 years later: applications and contributions from human-computer interaction},
	title = {Speed–accuracy tradeoff in {Fitts}’ law tasks—on the equivalency of actual and nominal pointing precision},
	volume = {61},
	issn = {1071-5819},
	url = {https://www.sciencedirect.com/science/article/pii/S1071581904001028},
	doi = {10.1016/j.ijhcs.2004.09.007},
	number = {6},
	urldate = {2025-04-09},
	journal = {International Journal of Human-Computer Studies},
	author = {Zhai, Shumin and Kong, Jing and Ren, Xiangshi},
	month = dec,
	year = {2004},
	pages = {823--856},
}

@inproceedings{qian_eyes_2017,
	address = {New York, NY, USA},
	series = {{SUI} '17},
	title = {The eyes don't have it: an empirical comparison of head-based and eye-based selection in virtual reality},
	isbn = {978-1-4503-5486-8},
	shorttitle = {The eyes don't have it},
	url = {https://dl.acm.org/doi/10.1145/3131277.3132182},
	doi = {10.1145/3131277.3132182},
	urldate = {2025-04-07},
	booktitle = {Proceedings of the 5th {Symposium} on {Spatial} {User} {Interaction}},
	publisher = {Association for Computing Machinery},
	author = {Qian, Yuan Yuan and Teather, Robert J.},
	month = oct,
	year = {2017},
	pages = {91--98},
}

@inproceedings{siddhpuria_pointing_2018,
	address = {New York, NY, USA},
	series = {{CHI} '18},
	title = {Pointing at a {Distance} with {Everyday} {Smart} {Devices}},
	isbn = {978-1-4503-5620-6},
	url = {https://dl.acm.org/doi/10.1145/3173574.3173747},
	doi = {10.1145/3173574.3173747},
	urldate = {2025-04-06},
	booktitle = {Proceedings of the 2018 {CHI} {Conference} on {Human} {Factors} in {Computing} {Systems}},
	publisher = {Association for Computing Machinery},
	author = {Siddhpuria, Shaishav and Malacria, Sylvain and Nancel, Mathieu and Lank, Edward},
	year = {2018},
	pages = {1--11},
}

@article{chowdhury_paws_2023,
	title = {{PAWS}: {Personalized} {Arm} and {Wrist} {Movements} {With} {Sensitivity} {Mappings} for {Controller}-{Free} {Locomotion} in {Virtual} {Reality}},
	volume = {7},
	shorttitle = {{PAWS}},
	url = {https://doi.org/10.1145/3604264},
	doi = {10.1145/3604264},
	number = {MHCI},
	urldate = {2025-04-09},
	journal = {Proc. ACM Hum.-Comput. Interact.},
	author = {Chowdhury, Sohan and Delamare, William and Irani, Pourang and Hasan, Khalad},
	month = sep,
	year = {2023},
	pages = {217:1--217:21},
}

@inproceedings{teather_pointing_2013,
	address = {New York, NY, USA},
	series = {{CHI} '13},
	title = {Pointing at 3d target projections with one-eyed and stereo cursors},
	isbn = {978-1-4503-1899-0},
	url = {https://dl.acm.org/doi/10.1145/2470654.2470677},
	doi = {10.1145/2470654.2470677},
	urldate = {2025-04-07},
	booktitle = {Proceedings of the {SIGCHI} {Conference} on {Human} {Factors} in {Computing} {Systems}},
	publisher = {Association for Computing Machinery},
	author = {Teather, Robert J. and Stuerzlinger, Wolfgang},
	year = {2013},
	pages = {159--168},
}

@inproceedings{lischke_magic-pointing_2016,
	address = {New York, NY, USA},
	series = {{CHI} {EA} '16},
	title = {{MAGIC}-{Pointing} on {Large} {High}-{Resolution} {Displays}},
	isbn = {978-1-4503-4082-3},
	url = {https://dl.acm.org/doi/10.1145/2851581.2892479},
	doi = {10.1145/2851581.2892479},
	urldate = {2025-04-06},
	booktitle = {Proceedings of the 2016 {CHI} {Conference} {Extended} {Abstracts} on {Human} {Factors} in {Computing} {Systems}},
	publisher = {Association for Computing Machinery},
	author = {Lischke, Lars and Schwind, Valentin and Friedrich, Kai and Schmidt, Albrecht and Henze, Niels},
	month = may,
	year = {2016},
	pages = {1706--1712},
}

@inproceedings{haque_myopoint_2015,
	address = {New York, NY, USA},
	series = {{CHI} '15},
	title = {Myopoint: {Pointing} and {Clicking} {Using} {Forearm} {Mounted} {Electromyography} and {Inertial} {Motion} {Sensors}},
	isbn = {978-1-4503-3145-6},
	shorttitle = {Myopoint},
	url = {https://dl.acm.org/doi/10.1145/2702123.2702133},
	doi = {10.1145/2702123.2702133},
	urldate = {2025-04-06},
	booktitle = {Proceedings of the 33rd {Annual} {ACM} {Conference} on {Human} {Factors} in {Computing} {Systems}},
	publisher = {Association for Computing Machinery},
	author = {Haque, Faizan and Nancel, Mathieu and Vogel, Daniel},
	year = {2015},
	pages = {3653--3656},
}

@inproceedings{nancel_high-precision_2013,
	address = {New York, NY, USA},
	series = {{CHI} '13},
	title = {High-precision pointing on large wall displays using small handheld devices},
	isbn = {978-1-4503-1899-0},
	url = {https://dl.acm.org/doi/10.1145/2470654.2470773},
	doi = {10.1145/2470654.2470773},
	urldate = {2025-03-25},
	booktitle = {Proceedings of the {SIGCHI} {Conference} on {Human} {Factors} in {Computing} {Systems}},
	publisher = {Association for Computing Machinery},
	author = {Nancel, Mathieu and Chapuis, Olivier and Pietriga, Emmanuel and Yang, Xing-Dong and Irani, Pourang P. and Beaudouin-Lafon, Michel},
	year = {2013},
	pages = {831--840},
}

@article{mackenzie_fitts_1992,
	title = {Fitts' {Law} as a {Research} and {Design} {Tool} in {Human}-{Computer} {Interaction}},
	volume = {7},
	issn = {0737-0024},
	url = {https://doi.org/10.1207/s15327051hci0701_3},
	doi = {10.1207/s15327051hci0701_3},
	number = {1},
	urldate = {2025-03-25},
	journal = {Human–Computer Interaction},
	author = {MacKenzie, I. Scott},
	month = mar,
	year = {1992},
	note = {Publisher: Taylor \& Francis},
	pages = {91--139},
}

@misc{crossman_speed_1957,
	title = {The speed and accuracy of hand movements},
	publisher = {Report to the MRC and DSIR Joint Committee on Individual  Efficiency in Industry},
	author = {Crossman, E.R.F.W.},
	year = {1957},
	note = {The Nature and Acquisition of Industrial Skill.},
  
}

@techreport{mine_virtual_1995,
	address = {USA},
	type = {Technical {Report}},
	title = {Virtual {Environment} {Interaction} {Techniques}},
	institution = {University of North Carolina at Chapel Hill},
	author = {Mine, Mark R.},
	year = {1995},
    url = {https://www.cs.unc.edu/techreports/95-018.pdf}
}

@article{schafer_controlling_2021,
	title = {Controlling {Teleportation}-{Based} {Locomotion} in {Virtual} {Reality} with {Hand} {Gestures}: {A} {Comparative} {Evaluation} of {Two}-{Handed} and {One}-{Handed} {Techniques}},
	volume = {10},
	copyright = {http://creativecommons.org/licenses/by/3.0/},
	issn = {2079-9292},
	shorttitle = {Controlling {Teleportation}-{Based} {Locomotion} in {Virtual} {Reality} with {Hand} {Gestures}},
	url = {https://www.mdpi.com/2079-9292/10/6/715},
	doi = {10.3390/electronics10060715},
	language = {en},
	number = {6},
	urldate = {2025-06-27},
	journal = {Electronics},
	author = {Schäfer, Alexander and Reis, Gerd and Stricker, Didier},
	month = jan,
	year = {2021},
	note = {Number: 6
Publisher: Multidisciplinary Digital Publishing Institute},
	pages = {715},
}

@inproceedings{kim_atatouch_2021,
	address = {New York, NY, USA},
	series = {{CHI} '21},
	title = {{AtaTouch}: {Robust} {Finger} {Pinch} {Detection} for a {VR} {Controller} {Using} {RF} {Return} {Loss}},
	isbn = {978-1-4503-8096-6},
	shorttitle = {{AtaTouch}},
	url = {https://dl.acm.org/doi/10.1145/3411764.3445442},
	doi = {10.1145/3411764.3445442},
	urldate = {2025-06-27},
	booktitle = {Proceedings of the 2021 {CHI} {Conference} on {Human} {Factors} in {Computing} {Systems}},
	publisher = {Association for Computing Machinery},
	author = {Kim, Daehwa and Park, Keunwoo and Lee, Geehyuk},
	month = may,
	year = {2021},
	pages = {1--9},
}

@article{bowman_pinch_2001,
	title = {Pinch keyboard: {Natural} text input for immersive virtual environments},
	shorttitle = {Pinch keyboard},
	author = {Bowman, Doug and Ly, Vinh and Campbell, Joshua and Tech, Virginia},
	month = jan,
	year = {2001},
    url = {http://hdl.handle.net/10919/20010}
}

@article{mangalam_enhancing_2024,
	title = {Enhancing hand-object interactions in virtual reality for precision manual tasks},
	volume = {28},
	issn = {1434-9957},
	url = {https://doi.org/10.1007/s10055-024-01055-3},
	doi = {10.1007/s10055-024-01055-3},
	language = {en},
	number = {4},
	urldate = {2025-06-27},
	journal = {Virtual Reality},
	author = {Mangalam, Madhur and Oruganti, Sanjay and Buckingham, Gavin and Borst, Christoph W.},
	month = nov,
	year = {2024},
	pages = {166},
}

@inproceedings{lee_comparison_2023,
	title = {Comparison of {Virtual} {Reality} {Teleportation} {Targeting} {Method} {Performance} depending on the {Teleport} {Distance}},
	url = {https://ieeexplore.ieee.org/document/10322242},
	doi = {10.1109/ISMAR-Adjunct60411.2023.00160},
	urldate = {2025-06-27},
	booktitle = {2023 {IEEE} {International} {Symposium} on {Mixed} and {Augmented} {Reality} {Adjunct} ({ISMAR}-{Adjunct})},
	author = {Lee, Jihyeon and Kim, Jinwook and Lee, Jeongmi},
	month = oct,
	year = {2023},
	note = {ISSN: 2771-1110},
	pages = {742--745},
}

@article{soukoreff_towards_2004,
	series = {Fitts' law 50 years later: applications and contributions from human-computer interaction},
	title = {Towards a standard for pointing device evaluation, perspectives on 27 years of {Fitts}’ law research in {HCI}},
	volume = {61},
	issn = {1071-5819},
	url = {https://www.sciencedirect.com/science/article/pii/S1071581904001016},
	doi = {10.1016/j.ijhcs.2004.09.001},
	number = {6},
	urldate = {2025-04-06},
	journal = {International Journal of Human-Computer Studies},
	author = {Soukoreff, R. William and MacKenzie, I. Scott},
	month = dec,
	year = {2004},
	pages = {751--789},
}

@inproceedings{triantafyllidis_challenges_2021,
	address = {New York, NY, USA},
	series = {{CHI} {EA} '21},
	title = {The {Challenges} in {Modeling} {Human} {Performance} in {3D} {Space} with {Fitts}’ {Law}},
	isbn = {978-1-4503-8095-9},
	url = {https://dl.acm.org/doi/10.1145/3411763.3443442},
	doi = {10.1145/3411763.3443442},
	urldate = {2025-03-25},
	booktitle = {Extended {Abstracts} of the 2021 {CHI} {Conference} on {Human} {Factors} in {Computing} {Systems}},
	publisher = {Association for Computing Machinery},
	author = {Triantafyllidis, Eleftherios and Li, Zhibin},
	month = may,
	year = {2021},
	pages = {1--9},
}

@article{kopper_human_2010,
	title = {A human motor behavior model for distal pointing tasks},
	volume = {68},
	issn = {1071-5819},
	url = {https://www.sciencedirect.com/science/article/pii/S1071581910000637},
	doi = {10.1016/j.ijhcs.2010.05.001},
	number = {10},
	urldate = {2025-03-25},
	journal = {International Journal of Human-Computer Studies},
	author = {Kopper, Regis and Bowman, Doug A. and Silva, Mara G. and McMahan, Ryan P.},
	month = oct,
	year = {2010},
	pages = {603--615},
}

@inproceedings{hourcade_how_2012,
	address = {New York, NY, USA},
	series = {{CHI} '12},
	title = {How small can you go? analyzing the effect of visual angle in pointing tasks},
	isbn = {978-1-4503-1015-4},
	shorttitle = {How small can you go?},
	url = {https://dl.acm.org/doi/10.1145/2207676.2207706},
	doi = {10.1145/2207676.2207706},
	urldate = {2025-04-06},
	booktitle = {Proceedings of the {SIGCHI} {Conference} on {Human} {Factors} in {Computing} {Systems}},
	publisher = {Association for Computing Machinery},
	author = {Hourcade, Juan Pablo and Bullock-Rest, Natasha},
	month = may,
	year = {2012},
	pages = {213--216},
}

@article{kovacs_perceptual_2008,
	title = {Perceptual influences on {Fitts}’ law},
	volume = {190},
	issn = {1432-1106},
	url = {https://doi.org/10.1007/s00221-008-1497-3},
	doi = {10.1007/s00221-008-1497-3},
	language = {en},
	number = {1},
	urldate = {2025-04-06},
	journal = {Experimental Brain Research},
	author = {Kovacs, A. J. and Buchanan, J. J. and Shea, C. H.},
	month = sep,
	year = {2008},
	pages = {99--103},
}

@misc{shi_pointing_2022,
	title = {Pointing {Cursor} {Interaction} in {Virtual} {Reality} from the {Perspective} of {Distance} {Perception} {\textbar} {IIETA}},
	url = {https://www.iieta.org/journals/ts/paper/10.18280/ts.390209},
	language = {en},
	urldate = {2025-04-06},
	author = {Shi, Mengdi and Hu, Tao and Yu, Jiawen},
	year = {2022},
	doi = {10.18280/ts.390209},
}

@inproceedings{janzen_modeling_2016,
	address = {New York, NY, USA},
	series = {{CHI} '16},
	title = {Modeling the {Impact} of {Depth} on {Pointing} {Performance}},
	isbn = {978-1-4503-3362-7},
	url = {https://dl.acm.org/doi/10.1145/2858036.2858244},
	doi = {10.1145/2858036.2858244},
	urldate = {2025-04-06},
	booktitle = {Proceedings of the 2016 {CHI} {Conference} on {Human} {Factors} in {Computing} {Systems}},
	publisher = {Association for Computing Machinery},
	author = {Janzen, Izabelle and Rajendran, Vasanth K. and Booth, Kellogg S.},
	month = may,
	year = {2016},
	pages = {188--199},
}

@article{tao_freehand_2021,
	title = {Freehand interaction with large displays: {Effects} of body posture, interaction distance and target size on task performance, perceived usability and workload},
	volume = {93},
	issn = {0003-6870},
	shorttitle = {Freehand interaction with large displays},
	url = {https://www.sciencedirect.com/science/article/pii/S000368702100017X},
	doi = {10.1016/j.apergo.2021.103370},
	urldate = {2025-04-06},
	journal = {Applied Ergonomics},
	author = {Tao, Da and Diao, Xiaofeng and Wang, Tieyan and Guo, Jingya and Qu, Xingda},
	month = may,
	year = {2021},
	pages = {103370},
}

@article{hoffman_vergenceaccommodation_2008,
	title = {Vergence–accommodation conflicts hinder visual performance and cause visual fatigue},
	volume = {8},
	issn = {1534-7362},
	url = {https://doi.org/10.1167/8.3.33},
	doi = {10.1167/8.3.33},
	number = {3},
	urldate = {2025-04-06},
	journal = {Journal of Vision},
	author = {Hoffman, David M. and Girshick, Ahna R. and Akeley, Kurt and Banks, Martin S.},
	month = mar,
	year = {2008},
	pages = {33},
}

@article{hart_nasa-task_2006,
	title = {Nasa-{Task} {Load} {Index} ({NASA}-{TLX}); 20 {Years} {Later}},
	volume = {50},
	issn = {1071-1813},
	url = {https://doi.org/10.1177/154193120605000909},
	doi = {10.1177/154193120605000909},
	language = {EN},
	number = {9},
	urldate = {2025-04-05},
	journal = {Proceedings of the Human Factors and Ergonomics Society Annual Meeting},
	author = {Hart, Sandra G.},
	month = oct,
	year = {2006},
	note = {Publisher: SAGE Publications Inc},
	pages = {904--908},
}

@inproceedings{barrera_machuca_effect_2019,
	address = {New York, NY, USA},
	series = {{CHI} '19},
	title = {The {Effect} of {Stereo} {Display} {Deficiencies} on {Virtual} {Hand} {Pointing}},
	isbn = {978-1-4503-5970-2},
	url = {https://dl.acm.org/doi/10.1145/3290605.3300437},
	doi = {10.1145/3290605.3300437},
	urldate = {2025-04-04},
	booktitle = {Proceedings of the 2019 {CHI} {Conference} on {Human} {Factors} in {Computing} {Systems}},
	publisher = {Association for Computing Machinery},
	author = {Barrera Machuca, Mayra Donaji and Stuerzlinger, Wolfgang},
	month = may,
	year = {2019},
	pages = {1--14},
}

@misc{noauthor_isots_2012,
	type = {International {Standard} confirmed [90.93]},
	title = {{ISO}/{TS} 9241-411:2012 {Ergonomics} of human-system interaction {Part} 411: {Evaluation} methods for the design of physical input devices},
	shorttitle = {{ISO}/{TS} 9241-411},
	url = {https://www.iso.org/standard/54106.html},
	language = {en},
	urldate = {2025-03-24},
	publisher = {ISO},
	month = may,
	year = {2012},
	note = {https://www.iso.org/standard/54106.html},
}

@article{westermeier_assessing_2024,
	title = {Assessing {Depth} {Perception} in {VR} and {Video} {See}-{Through} {AR}: {A} {Comparison} on {Distance} {Judgment}, {Performance}, and {Preference}},
	volume = {30},
	issn = {1941-0506},
	shorttitle = {Assessing {Depth} {Perception} in {VR} and {Video} {See}-{Through} {AR}},
	url = {https://ieeexplore.ieee.org/document/10458408},
	doi = {10.1109/TVCG.2024.3372061},
	number = {5},
	urldate = {2025-03-26},
	journal = {IEEE Transactions on Visualization and Computer Graphics},
	author = {Westermeier, Franziska and Brübach, Larissa and Wienrich, Carolin and Latoschik, Marc Erich},
	month = may,
	year = {2024},
	note = {Conference Name: IEEE Transactions on Visualization and Computer Graphics},
	pages = {2140--2150},
}

@inproceedings{xu_pointing_2019,
	title = {Pointing and {Selection} {Methods} for {Text} {Entry} in {Augmented} {Reality} {Head} {Mounted} {Displays}},
	url = {https://ieeexplore.ieee.org/document/8943748},
	doi = {10.1109/ISMAR.2019.00026},
	urldate = {2025-03-25},
	booktitle = {2019 {IEEE} {International} {Symposium} on {Mixed} and {Augmented} {Reality} ({ISMAR})},
	author = {Xu, Wenge and Liang, Hai-Ning and He, Anqi and Wang, Zifan},
	month = oct,
	year = {2019},
	note = {ISSN: 1554-7868},
	pages = {279--288},
}

@inproceedings{zhang_double_2017,
	title = {Double hand-gesture interaction for walk-through in {VR} environment},
	url = {https://ieeexplore.ieee.org/document/7960051},
	doi = {10.1109/ICIS.2017.7960051},
	urldate = {2025-03-25},
	booktitle = {2017 {IEEE}/{ACIS} 16th {International} {Conference} on {Computer} and {Information} {Science} ({ICIS})},
	author = {Zhang, Fan and Chu, Shaowei and Pan, Ruifang and Ji, Naye and Xi, Lian},
	month = may,
	year = {2017},
	pages = {539--544},
}

@inproceedings{sindhupathiraja_exploring_2024,
	title = {Exploring {Bi}-{Manual} {Teleportation} in {Virtual} {Reality}},
	url = {https://ieeexplore.ieee.org/document/10494103},
	doi = {10.1109/VR58804.2024.00095},
	urldate = {2025-03-25},
	booktitle = {2024 {IEEE} {Conference} {Virtual} {Reality} and {3D} {User} {Interfaces} ({VR})},
	author = {Sindhupathiraja, Siddhanth Raja and Ullah, A K M Amanat and Delamare, William and Hasan, Khalad},
	month = mar,
	year = {2024},
	note = {ISSN: 2642-5254},
	pages = {754--764},
}

@inproceedings{schafer_controlling_2022,
	address = {Cham},
	title = {Controlling {Continuous} {Locomotion} in {Virtual} {Reality} with {Bare} {Hands} {Using} {Hand} {Gestures}},
	isbn = {978-3-031-16234-3},
	doi = {10.1007/978-3-031-16234-3_11},
	language = {en},
	booktitle = {Virtual {Reality} and {Mixed} {Reality}},
	publisher = {Springer International Publishing},
	author = {Schäfer, Alexander and Reis, Gerd and Stricker, Didier},
	editor = {Zachmann, Gabriel and Alcañiz Raya, Mariano and Bourdot, Patrick and Marchal, Maud and Stefanucci, Jeanine and Yang, Xubo},
	year = {2022},
	pages = {191--205},
}

@inproceedings{prithul_evaluation_2022,
	address = {New York, NY, USA},
	series = {{SUI} '22},
	title = {Evaluation of {Hands}-free {Teleportation} in {VR}},
	isbn = {978-1-4503-9948-7},
	url = {https://dl.acm.org/doi/10.1145/3565970.3567683},
	doi = {10.1145/3565970.3567683},
	urldate = {2025-03-25},
	booktitle = {Proceedings of the 2022 {ACM} {Symposium} on {Spatial} {User} {Interaction}},
	publisher = {Association for Computing Machinery},
	author = {Prithul, Aniruddha and Bhandari, Jiwan and Spurgeon, Walker and Folmer, Eelke},
	year = {2022},
	pages = {1--6},
}

@inproceedings{oakley_pointing_2008,
	address = {New York, NY, USA},
	series = {{CHI} {EA} '08},
	title = {Pointing with fingers, hands and arms for wearable computing},
	isbn = {978-1-60558-012-8},
	url = {https://dl.acm.org/doi/10.1145/1358628.1358840},
	doi = {10.1145/1358628.1358840},
	urldate = {2025-03-25},
	booktitle = {{CHI} '08 {Extended} {Abstracts} on {Human} {Factors} in {Computing} {Systems}},
	publisher = {Association for Computing Machinery},
	author = {Oakley, Ian and Sunwoo, John and Cho, Il-Yeon},
	year = {2008},
	pages = {3255--3260},
}

@article{naceri_depth_2010,
	title = {Depth {Perception} {Within} {Virtual} {Environments}: {Comparison} {Between} two {Display} {Technologies}},
	volume = {3},
	shorttitle = {Depth {Perception} {Within} {Virtual} {Environments}},
	journal = {International Journal on Advances in Intelligent Systems},
	author = {Naceri, Djallil and Chellali, Ryad and Dionnet, Fabien and Toma, Simone},
	month = jan,
	year = {2010},
	pages = {51--64},
    url = {https://www.researchgate.net/publication/242330470} 
}

@inproceedings{myers_interacting_2002,
	address = {New York, NY, USA},
	series = {{CHI} '02},
	title = {Interacting at a distance: measuring the performance of laser pointers and other devices},
	isbn = {978-1-58113-453-7},
	shorttitle = {Interacting at a distance},
	url = {https://dl.acm.org/doi/10.1145/503376.503383},
	doi = {10.1145/503376.503383},
	urldate = {2025-03-25},
	booktitle = {Proceedings of the {SIGCHI} {Conference} on {Human} {Factors} in {Computing} {Systems}},
	publisher = {Association for Computing Machinery},
	author = {Myers, Brad A. and Bhatnagar, Rishi and Nichols, Jeffrey and Peck, Choon Hong and Kong, Dave and Miller, Robert and Long, A. Chris},
	year = {2002},
	pages = {33--40},
}

@article{monteiro_evaluation_2023,
	title = {Evaluation of {Hands}-{Free} {VR} {Interaction} {Methods} {During} a {Fitts}’ {Task}: {Efficiency} and {Effectiveness}},
	volume = {11},
	issn = {2169-3536},
	shorttitle = {Evaluation of {Hands}-{Free} {VR} {Interaction} {Methods} {During} a {Fitts}’ {Task}},
	url = {https://ieeexplore.ieee.org/document/10176123},
	doi = {10.1109/ACCESS.2023.3293057},
	urldate = {2025-03-25},
	journal = {IEEE Access},
	author = {Monteiro, Pedro and Gonçalves, Guilherme and Peixoto, Bruno and Melo, Miguel and Bessa, Maximino},
	year = {2023},
	note = {Conference Name: IEEE Access},
	pages = {70898--70911},
}

@article{lin_investigation_2015,
	title = {An investigation of pointing postures in a {3D} stereoscopic environment},
	volume = {48},
	issn = {0003-6870},
	url = {https://www.sciencedirect.com/science/article/pii/S0003687014002919},
	doi = {10.1016/j.apergo.2014.12.001},
	urldate = {2025-03-25},
	journal = {Applied Ergonomics},
	author = {Lin, Chiuhsiang Joe and Ho, Sui-Hua and Chen, Yan-Jyun},
	month = may,
	year = {2015},
	pages = {154--163},
}

@inproceedings{kyto_pinpointing_2018,
	address = {New York, NY, USA},
	series = {{CHI} '18},
	title = {Pinpointing: {Precise} {Head}- and {Eye}-{Based} {Target} {Selection} for {Augmented} {Reality}},
	isbn = {978-1-4503-5620-6},
	shorttitle = {Pinpointing},
	url = {https://dl.acm.org/doi/10.1145/3173574.3173655},
	doi = {10.1145/3173574.3173655},
	urldate = {2025-03-25},
	booktitle = {Proceedings of the 2018 {CHI} {Conference} on {Human} {Factors} in {Computing} {Systems}},
	publisher = {Association for Computing Machinery},
	author = {Kytö, Mikko and Ens, Barrett and Piumsomboon, Thammathip and Lee, Gun A. and Billinghurst, Mark},
	year = {2018},
	pages = {1--14},
}

@inproceedings{khundam_first_2015,
	title = {First person movement control with palm normal and hand gesture interaction in virtual reality},
	url = {https://ieeexplore.ieee.org/document/7219818},
	doi = {10.1109/JCSSE.2015.7219818},
	urldate = {2025-03-25},
	booktitle = {2015 12th {International} {Joint} {Conference} on {Computer} {Science} and {Software} {Engineering} ({JCSSE})},
	author = {Khundam, Chaowanan},
	month = jul,
	year = {2015},
	pages = {325--330},
}

@inproceedings{jones_effects_2008,
	address = {New York, NY, USA},
	series = {{APGV} '08},
	title = {The effects of virtual reality, augmented reality, and motion parallax on egocentric depth perception},
	isbn = {978-1-59593-981-4},
	url = {https://dl.acm.org/doi/10.1145/1394281.1394283},
	doi = {10.1145/1394281.1394283},
	urldate = {2025-03-25},
	booktitle = {Proceedings of the 5th symposium on {Applied} perception in graphics and visualization},
	publisher = {Association for Computing Machinery},
	author = {Jones, J. Adam and Swan, J. Edward and Singh, Gurjot and Kolstad, Eric and Ellis, Stephen R.},
	year = {2008},
	pages = {9--14},
}

@article{huang_design_2019,
	title = {Design of finger gestures for locomotion in virtual reality},
	volume = {1},
	issn = {2096-5796},
	url = {https://www.sciencedirect.com/science/article/pii/S2096579619300026},
	doi = {10.3724/SP.J.2096-5796.2018.0007},
	number = {1},
	urldate = {2025-03-24},
	journal = {Virtual Reality \& Intelligent Hardware},
	author = {Huang, Rachel and Harris-adamson, Carisa and Odell, Dan and Rempel, David},
	month = feb,
	year = {2019},
	pages = {1--9},
}

@inproceedings{hincapie-ramos_consumed_2014,
	address = {New York, NY, USA},
	series = {{CHI} '14},
	title = {Consumed endurance: a metric to quantify arm fatigue of mid-air interactions},
	isbn = {978-1-4503-2473-1},
	shorttitle = {Consumed endurance},
	url = {https://dl.acm.org/doi/10.1145/2556288.2557130},
	doi = {10.1145/2556288.2557130},
	urldate = {2025-03-24},
	booktitle = {Proceedings of the {SIGCHI} {Conference} on {Human} {Factors} in {Computing} {Systems}},
	publisher = {Association for Computing Machinery},
	author = {Hincapié-Ramos, Juan David and Guo, Xiang and Moghadasian, Paymahn and Irani, Pourang},
	year = {2014},
	pages = {1063--1072},
}

@inproceedings{hansen_fitts_2018,
	address = {New York, NY, USA},
	series = {{COGAIN} '18},
	title = {A {Fitts}' law study of click and dwell interaction by gaze, head and mouse with a head-mounted display},
	isbn = {978-1-4503-5790-6},
	url = {https://dl.acm.org/doi/10.1145/3206343.3206344},
	doi = {10.1145/3206343.3206344},
	urldate = {2025-03-24},
	booktitle = {Proceedings of the {Workshop} on {Communication} by {Gaze} {Interaction}},
	publisher = {Association for Computing Machinery},
	author = {Hansen, John Paulin and Rajanna, Vijay and MacKenzie, I. Scott and Bækgaard, Per},
	month = jun,
	year = {2018},
	pages = {1--5},
}

@inproceedings{franzluebbers_versatile_2023,
	address = {New York, NY, USA},
	series = {{VRST} '23},
	title = {Versatile {Mixed}-method {Locomotion} under {Free}-hand and {Controller}-based {Virtual} {Reality} {Interfaces}},
	isbn = {979-8-4007-0328-7},
	url = {https://dl.acm.org/doi/10.1145/3611659.3615701},
	doi = {10.1145/3611659.3615701},
	urldate = {2025-03-24},
	booktitle = {Proceedings of the 29th {ACM} {Symposium} on {Virtual} {Reality} {Software} and {Technology}},
	publisher = {Association for Computing Machinery},
	author = {Franzluebbers, Anton and Johnsen, Kyle},
	month = oct,
	year = {2023},
	pages = {1--10},
}

@inproceedings{czerwinski_toward_2003,
	title = {Toward {Characterizing} the {Productivity} {Benefits} of {Very} {Large} {Displays}},
	url = {https://www.semanticscholar.org/paper/Toward-Characterizing-the-Productivity-Benefits-of-Czerwinski-Smith/d3076f1e3c19bcf2a46922294aa67ddb1d36b34b},
	urldate = {2025-03-24},
	author = {Czerwinski, M. and Smith, Greg and Regan, Tim and Meyers, B. and Robertson, G. and Starkweather, G.},
	year = {2003},
}

@article{cao_real-time_2023,
	title = {Real-time multimodal interaction in virtual reality - a case study with a large virtual interface},
	volume = {82},
	issn = {1380-7501},
	url = {https://doi.org/10.1007/s11042-023-14381-6},
	doi = {10.1007/s11042-023-14381-6},
	number = {16},
	urldate = {2025-03-24},
	journal = {Multimedia Tools Appl.},
	author = {Cao, Lizhou and Zhang, Huadong and Peng, Chao and Hansberger, Jeffrey T.},
	month = feb,
	year = {2023},
	pages = {25427--25448},
}

@book{laviola_3d_2017,

	edition = {Second edition},
	series = {Pearson always learning},
	title = {{3D} user interfaces: theory and practice},
	isbn = {978-0-13-403432-4},
	shorttitle = {{3D} user interfaces},
	language = {eng},
	publisher = {Addison-Wesley},
	author = {LaViola, Joseph J. and Kruijff, Ernst and McMahan, Ryan P. and Bowman, Doug A. and Poupyrev, Ivan},
	year = {2017},
}

@article{bernardos_comparison_2016,
	title = {A {Comparison} of {Head} {Pose} and {Deictic} {Pointing} {Interaction} {Methods} for {Smart} {Environments}},
	volume = {32},
	issn = {1044-7318},
	url = {https://www.tandfonline.com/doi/full/10.1080/10447318.2016.1142054},
	doi = {10.1080/10447318.2016.1142054},
	number = {4},
	urldate = {2025-03-24},
	journal = {International Journal of Human–Computer Interaction},
	author = {Bernardos, Ana M. and , David, Gómez and and Casar, José R.},
	month = apr,
	year = {2016},
	note = {Publisher: Taylor \& Francis},
	pages = {325--351},
}

@inproceedings{batmaz_head-mounted_2019,
	title = {Do {Head}-{Mounted} {Display} {Stereo} {Deficiencies} {Affect} {3D} {Pointing} {Tasks} in {AR} and {VR}?},
	url = {https://ieeexplore.ieee.org/document/8797975},
	doi = {10.1109/VR.2019.8797975},
	urldate = {2025-03-24},
	booktitle = {2019 {IEEE} {Conference} on {Virtual} {Reality} and {3D} {User} {Interfaces} ({VR})},
	author = {Batmaz, Anil Ufuk and Machuca, Mayra Donaji Barrera and Pham, Duc Minh and Stuerzlinger, Wolfgang},
	month = mar,
	year = {2019},
	note = {ISSN: 2642-5254},
	pages = {585--592},
}

@article{argelaguet_efficient_2009,
	title = {Efficient {3D} {Pointing} {Selection} in {Cluttered} {Virtual} {Environments}},
	volume = {29},
	issn = {1558-1756},
	url = {https://ieeexplore.ieee.org/document/5307641},
	doi = {10.1109/MCG.2009.117},
	number = {6},
	urldate = {2025-03-24},
	journal = {IEEE Computer Graphics and Applications},
	author = {Argelaguet, Ferran and Andujar, Carlos},
	month = nov,
	year = {2009},
	note = {Conference Name: IEEE Computer Graphics and Applications},
	pages = {34--43},
}
\end{document}